\documentclass[preprints,article,accept,moreauthors,pdftex]{Definitions/mdpi} 
\usepackage{mathtools} 
\usepackage{amsmath}
\usepackage{amsfonts}
\usepackage{amssymb}

\renewcommand{\BibitemShut}[1]{}

\newcommand{\bi}{\begin{itemize}} 		\newcommand{\ei}{\end{itemize}}
\newcommand{\benu}{\begin{enumerate}} \newcommand{\enu}{\end{enumerate}}
\newcommand{\bd}{\begin{dinglist}{0}}     \newcommand{\ed}{\end{dinglist}}
\newcommand{\bfig}{\begin{figure}[htbp]}  \newcommand{\efig}{\end{figure}}
       			
\newcommand{\bc}{\begin{center}} 	\newcommand{\ec}{\end{center}}
\newcommand{\be}{\begin{equation}} 	\newcommand{\ee}{\end{equation}}
\newcommand{\bsub}{\begin{subequations}}  \newcommand{\esub}{\end{subequations}}
\newcommand{\ben}{\begin{eqnarray}} 	\newcommand{\een}{\end{eqnarray}}
\newcommand{\ba}[1]{\begin{array}{#1}} 	\newcommand{\ea}{\end{array}}
\newcommand{\bea}{\be\begin{array}{rcl}} \newcommand{\eea}{\end{array}\ee}

\usepackage{verbatim}
\newcommand{\HCd}{\mathcal{H}}

\def\HCdt0{\tilde{\HCd}_{0}}

\newcommand\br{\begin{eqnarray}}
\newcommand\er{\end{eqnarray}}

\renewcommand\({\left(}
\renewcommand\){\right)}

\renewcommand\a{\alpha}

\renewcommand\d{\delta}

\renewcommand\G{\Gamma}

\renewcommand\k{\kappa}
\renewcommand\l{\lambda}

\newcommand\m{\mu}
\newcommand\n{\nu}

\renewcommand\O{\Omega}
\newcommand\p{\phi}

\renewcommand\P{\Phi}

\renewcommand\r{\rho}

\renewcommand\S{\Sigma}

\newcommand\twomat[4]{\left(\begD{array}{cc}  
{#1} & {#2} \\ {#3} & {#4} \end{array} \right)}

\definecolor{red}{rgb}{1,0,0}

\def\p{\partial}

\def\+{^\dagger}

\def\<{\leftarrow}
\def\>{\rightarrow}
\def\({\left(}
\def\){\right)}

\def\P{{\cal P}}\def\S{{\cal S}}

\def\a{\alpha}   \def\d{\delta} \def\e{\epsilon}
\def\m{\mu} \def\n{\nu} \def\r{\rho}  \def\l{\lambda} 
\def\k{\kappa}\def\G{\Gamma}\def\O{\Omega}

\newcommand{\LL}{\mathcal{L}}

\firstpage{1} 
\pubvolume{xx}
\issuenum{1}
\articlenumber{5}
\pubyear{2019}
\copyrightyear{2019}
\history{Received: date; Accepted: date; Published: date}

\Title{Singularity-free and cosmologically viable Born-Infeld gravity with scalar matter}

\Author{David Benisty$^{1,2}$, Gonzalo J. Olmo$^{3}$, Diego Rubiera-Garcia$^{* 4}$}

\address{%
$^{1}$ \quad DAMTP, Centre for Mathematical Sciences, University of Cambridge, Wilberforce Road, Cambridge CB3 0WA, United Kingdom \\
$^{2}$ \quad Kavli Institute of Cosmology (KICC), University of Cambridge, Madingley Road, Cambridge, CB3 0HA, UK \\
$^{3}$ \quad Departamento de F\'{i}sica Te\'{o}rica and IFIC, Centro Mixto Universidad de Valencia - CSIC.
Universidad de Valencia, Burjassot-46100, Valencia, Spain\\
$^{4}$ \quad Departamento de F\'isica Te\'orica and IPARCOS, Universidad Complutense de Madrid, E-28040
Madrid, Spain\\}

\corres{Correspondence: drubiera@ucm.es}

\abstract{The early Cosmology driven by a single scalar field, both massless and massive, in the context of Eddington-inspired Born-Infeld gravity, is explored. We show the existence of nonsingular solutions of bouncing and loitering type (depending on the sign of the gravitational theory's parameter, $\epsilon$) replacing the Big Bang singularity, and discuss their properties. In addition, in the massive case we find some new features of the cosmological evolution depending on the value of the mass parameter, including asymmetries in the expansion/contraction phases, or a continuous transition between a contracting phase to an expanding one via an intermediate loitering phase. We also provide a combined analysis of cosmic chronometers, standard candles, BAO, and CMB data to constrain the model, finding that for roughly  $\vert \epsilon \vert \lesssim 5\cdot 10^{-8} \text{ m}^2$ the model is compatible with the latest observations while successfully removing the Big Bang singularity. This bound is several orders of magnitude stronger than the most stringent constraints currently available in the literature.}

\begin{document}








\section{Introduction} \label{sec:I}

The standard concordance cosmological $\Lambda$CDM model, framed within Einstein's General Theory of Relativity (GR), including an early phase of inflationary expansion, a cold dark matter component, and a tiny cosmological constant driving the accelerated late-time expansion of the Universe, has successfully met all observations \cite{Amendola:2012ys,Aghanim:2018eyx}. Within this model, scalar fields have found new and imaginative applications. For instance,  inflationary models in the early Universe involve from one to many scalar fields \cite{Starobinsky:1979ty,Starobinsky:1980te,Guth:1980zm,Albrecht:1982wi,Mukhanov:1981xt,Guth:1982ec,Linde:1981mu,Barrow:1988xh,Barrow:1988xi,Elizalde:2008yf}. In the slow-roll approximation the exact form of the scalar field potential is unknown since many different potentials have been studied and confronted to observations. On the other hand, a way to parameterize dark energy is by using a scalar field, the so-called quintessence model (for canonical scalar fields \cite{Ratra:1987rm,Caldwell:1997ii}) or its generalizations to K-essence models (when a non-canonical scalar Lagrangian is considered \cite{Kehayias:2019gir,Oikonomou:2019muq,Chakraborty:2019swx,Babichev:2018twg}), in such a way that the cosmological constant gets replaced by a dark energy fluid with a nearly constant density today \cite{Zlatev:1998tr,Caldwell:1999ew,Chiba:1999ka,Bento:2002ps,Tsujikawa:2013fta}. Dark matter can be also parameterized in terms of weakly-interacting massive particles, which can be scalar particles still undiscovered at colliders and other dark matter detection experiments. Models for dark matter can also be based on other kinds of scalar fields, for instance, via fuzzy dark matter \cite{Hu:2000ke}, or by using a Lagrange multiplier that changes the behaviour of the kinetic term \cite{Anagnostopoulos:2019myt,Benisty:2018oyy,Benisty:2018qed,Benisty:2017eqh}. Scalar field models may also be the result of complex effective interactions of other fundamental fields in equilibrium, such as in Bose-Einstein condensates, thus allowing for an even broader range of phenomenological justifications.  Therefore, the consideration of scalar fields within cosmological models is justified from the point of view of building a fully consistent history of the cosmological evolution compatible with observations.

On the gravity side, and despite its observational success, from a fundamental point of view the $\Lambda$CDM model still contains a visible singularity at the Universe's past. This is an unavoidable consequence of the singularity theorems (see e.g. \cite{Senovilla:2014gza} for a pedagogical discussion). In order to tackle this issue, it is widely assumed in the community that at the strong curvatures and fields of the very early Universe, quantum gravity effects should come into play in order to regularize this singularity. Since a quantum theory of gravity is not available yet, an effective way to capture such hypothetical effects  is via modified theories of gravity \cite{DeFelice:2010aj,CLreview,Nojiri:2017ncd,Heisenberg:2018vsk}. The plausibility of such alternative gravitational descriptions to supersede GR and represent observationally viable alternatives to the $\Lambda$CDM model has been widely discussed in the literature according to different perspectives and approaches  \cite{Bull:2015stt}.

Among the large pool of theories which have been  investigated in the literature, for the sake of this paper we bring here the proposal originally introduced by Banados and Ferreira \cite{banados} and dubbed as Eddington-inspired Born-Infeld (EiBI) gravity\footnote{This proposal can actually be framed within the tradition of considering square-root action, such as in the DBI one, see e.g. \cite{Alishahiha:2004eh,Liu:2012rc,Choudhury:2012yh,Choudhury:2015yna}.}. In order to avoid troubles with ghost-like instabilities, this theory is typically formulated in metric-affine spaces, where metric and affine connection are a priori independent entities. EiBI theory has found many different applications in astrophysics and cosmology, see e.g. \cite{Harko:2013wka,Wei:2014dka,Shaikh:2015oha,Avelino:2015fve,Prasetyo:2017hrb,Chen:2017ify,Shaikh:2018yku,Jana:2018knq,Boehmer:2019uxv,Delhom:2019btt}. In particular, the existence of bouncing solutions replacing the Big Bang singularity within EiBI gravity was first hinted in \cite{Avelino:2012ue}, in particular, when scalar fields are considered as the matter source. Further works in the subject in the last few years have reinforced the capability of this theory to remove such singularities according to different mechanisms (for a review, see \cite{BeltranJimenez:2017doy}).

The main aim of this work is to construct explicit such singularity-free solutions within the early cosmological evolution, corresponding to a single massless (quintaessential) scalar field, and to further extend (numerically) this analysis to the massive case. We shall show the existence of two kinds of singularity-free solutions depending on the sign of the EiBI gravity parameter. The first one corresponds to bouncing solutions, where the universe contracts down to a minimum size before entering into an expansion phase, while the second are loitering solutions, which interpolate between an asymptotically Minkowski past and the current cosmological evolution. For these nonsingular solutions we carry out a combined analysis of cosmic chronometers, standard candles, BAO, and CMB data in order to constrain the EiBI parameter, finding the bound $\vert \epsilon \vert \lesssim 5\cdot 10^{-8} \text{ m}^2$.

This paper is organized as follows: in Sec.\ref{sec:II} we introduce EiBI gravity, discuss its properties, and construct its cosmological equations when coupled to a scalar field.  In Sec.\ref{sec:III} we consider massless scalar fields and discuss its corresponding bouncing and loitering solutions, extending these results in Sec.\ref{sec:IV} to the massive case. Sec.\ref{sec:obs} sees the theory fit with the latest observations (on $\Lambda$CDM background), and we conclude in Sec.\ref{sec:V} with a summary and some perspectives.

\section{Eddington-inspired Born-Infeld gravity} \label{sec:II}

\subsection{Action and basic field equations}

The action of EiBI gravity can be conveniently written as
\be \label{eq:SBIm}
\mathcal{S}_{EiBI}=\frac{1}{\e\k^2} \int d^4x \left[ \sqrt{-q } - \l \sqrt{-g } \right] + \mathcal{S}_m(g_{\m\n},\psi_m) \ ,
\ee
where
$\k^2 \equiv 8\pi G/c^4$ is Newton's constant, $\e$ is EiBI parameter, $\lambda$ is a dimensionless constant, $g$ is the determinant of the space-time metric
 $g_{\m\n}$ and $q$ the determinant of an auxiliary metric defined as:
 \begin{equation}
 q_{\m\n}\equiv g_{\m\n} + \e R_{(\m\n)}(\G)    \ ,
 \end{equation}
 where the (symmetric part of the) Ricci tensor $R_{\m\n}(\G) \equiv {R^\r}_{\m\r\n}(\G)$  is a function solely of the (torsionless) affine connection $\G \equiv \G_{\m\n}^{\l}$, assumed to  be {\it a priori} independent of the space-time metric $g_{\m\n}$ (metric-affine or Palatini formalism). This symmetrizing requirement ensures that the theory is invariant under projective transformations, which avoids the presence of ghost-like instabilities \cite{BeltranJimenez:2019acz,Jimenez:2020dpn}. As for the matter sector, $\mathcal{S}_m=\int d^4x \sqrt{-g} \mathcal{L}_m(g_{\mu\nu},\psi_m)$, it is assumed to be minimally coupled to the space-time metric $g_{\m\n}$ (see e.g. \cite{Delhom:2020hkb} for a definition of minimal coupling in metric-affine theories), with $\psi_m$ denoting collectively the matter fields. It is worth pointing out that EiBI gravity recovers the GR dynamics and its solutions in the $\vert R_{\mu\nu} \vert  \ll \epsilon^{-1}$ limit \cite{BeltranJimenez:2017doy} (from now on, vertical bars indicate a determinant), inheriting an effective cosmological constant $ \Lambda_{eff}=\frac{\lambda-1}{\epsilon \kappa^2}$. This fact allows EiBI gravity to naturally pass weak-field limit tests, such as those based on solar system observations.

EiBI gravity is nowadays a very well known theory in the community thanks to its many applications (for a detailed account of its properties we refer the reader to the review \cite{BeltranJimenez:2017doy}). It is actually a member of the so-called Ricci-based family of gravitational theories \cite{Afonso:2018bpv}, all of which admit an Einstein-like representation of their field equations given by
\be \label{eq:Rmunuq}
{G^\m}_{\n}(q)=\frac{\k^2}{| \O |^{1/2}} \left[{T^\m}_{\n} - \delta^{\m}_{\n}\left( \mathcal{L}_{G} + \frac{T}{2}\right) \right] \ ,
\ee
where $\mathcal{L}_G$ is the gravitational Lagrangian, $T_{\m\n} = \frac{2}{\sqrt{-g}} \frac{\d S_m}{\d g^{\m\n}}$ is the stress-energy tensor of the matter fields, $T$ its trace, and ${G^\m}_{\n}(q)$ is the Einstein tensor of the auxiliary metric $q_{\mu\nu}$, such that $\Gamma$ is Levi-Civita of it, that is
\begin{equation}
\nabla_{\a} (\sqrt{-q} \, q^{\m\n})=0 \ .
\end{equation}
This $q_{\mu\nu}$ metric is related to the space-time one $g_{\mu\nu}$ via the relation
\be \label{eq:qg}
q_{\m\n} =  g_{\m\a}\, {\O^\a}_\n \ ,
\ee
where the \emph{deformation} matrix $\hat{\Omega}$ depends on-shell on the matter fields (and possibly the space-time metric $g_{\mu\nu}$ as well). For EiBI gravity this matrix is implicitly determined via the equation
\be \label{eq:veromega}
|\O|^{1/2}{(\O^{-1})^\m}_\n=\l \d^{\m}_{\n}  -\e\k^2{T^\m}_\n \ ,
\ee
while the EiBI Lagrangian in (\ref{eq:SBIm}) can also be expressed in terms of this matrix as
\be
\mathcal{L}_G=\frac{ | \O |^{1/2} - \l}{\e \k^2} \ .
\ee
Let us point out that all terms on the right-hand side of the equations (\ref{eq:Rmunuq}) are functions of the matter fields and the metric $g_{\mu\nu}$, thus representing a system of second-order field equations with new couplings engendered by the matter fields. In vacuum, ${T^\mu}_{\nu}=0$, one recovers the GR solutions\footnote{As a remark, we would like to note that it has been shown that in generic RBGs, due to the non-linearities of the field equations, the deformation matrix admits other (typically pathological) solutions besided the one that boils down to GR in vacuum. These solutions can lead to anisotropic deformation matrices even if the stress-energy tensor is isotropic. Remarkably, it was shown that for EiBI gravity, no such anisotropic solutions exist in presence of an isotropic stress-energy tensor, see \cite{Jimenez:2020iok}.}, which ensures the propagation of the two polarizations of the gravitational field travelling at the speed of light \cite{Jana:2017ost}.

\subsection{EiBI cosmology with scalar fields}

As the matter sector of our model let us consider a single (real) scalar field described by the action
\be \label{eq:matteraction}
\mathcal{S}_m=-\frac{1}{2}\int d^4x \sqrt{-g} \LL_m=-\frac{1}{2}\int d^4x \sqrt{-g} (X+2V(\phi)) \ ,
\ee
with kinetic term $X\equiv g^{\m\n}\p_\m\phi\, \p_\n\phi$  and scalar potential $V(\phi)$. The variation of this action with respect to the scalar field leads to the field equations
\be\label{eq:phi}
\frac{1}{\sqrt{-g}}\p_\m\(\sqrt{-g} g^{\m\n}\p_\n \phi\) = V_\phi \ ,
\ee
(where $V_{\phi} \equiv dV/d\phi$) and to the associated stress-energy tensor
\be\label{eq:Tmn}
{T^\m}_{\n}=g^{\m\a}\p_{\a}\phi\p_{\n}\phi-\frac{\LL_m}{2}\d^{\m}_{\n} \ .
\ee
We consider next a spatially flat Friedman-Lemaitre-Robertson-Walker (FLRW) space-time\footnote{For simplicity in this work we do not consider the cases of open and closed universes, though the analysis could also be carried out in such cases following, for instance, the results of \cite{Barragan:2010qb}.}, with line element:
\be\label{ds2}
ds_g^2 \equiv g_{\m\n}dx^{\m}dx^{\n} =-dt^2+a^2(t) d\vec{x}^2 \ ,
\ee
where $a(t)$ is the expansion factor and $ d\vec{x}^{\,2}=\d_{ij} \,dx^{i}dx^{j}$, with $i, j =1\ldots3$ for the spatial part. As from now on we shall assume $\phi=\phi(t)$,
the stress-energy tensor \eqref{eq:Tmn} reads:
\be  \label{eq:Tttij}
{T^\m}_\n=\(\begin{array}{cc} -\tfrac12\dot{\phi}^2 - V  & 0 \\ 0 & (\tfrac12\dot{\phi}^2 - V) {\bf\it I}_{3\times3} \end{array}\) \ .
\ee
where a dot denotes a derivative with respect to $t$, as usual.

From Eq.\eqref{eq:veromega} we can conclude that the deformation matrix needs to have a similar algebraic (diagonal) structure as that of the stress-energy tensor of the scalar field, namely
\be \label{eq:omcos}
{\O^\m}_\n=\(\begin{array}{cc}  \O_+ & 0 \\ 0 & \O_- \,{\bf\it I}_{3\times3} \end{array}\) \ ,
\ee
where the identity matrix $I_{3\times 3}$ represents the spatial sector. Now, plugging this ansatz and Eq.(\ref{eq:Tttij}) into Eq.(\ref{eq:veromega}) gives the components of this matrix as:
\ben
\O_-^2  &=&   \left(\l+\e\k^2V -\tfrac{\e\k^2\dot{\phi} ^2}{2}\right)\left(\l+\e\k^2V +\tfrac{\e\k^2\dot{\phi} ^2}{2}\right)\nonumber \\ &=& \left(\tilde \lambda^2-\Phi^2\right)\label{Om}\\
\O_+^2  &=&   \frac{\left(\l+\e\k^2V -\tfrac{\e\k^2\dot{\phi} ^2}{2}\right)^{3}}{\left(\l+\e\k^2V +\tfrac{\e\k^2\dot{\phi} ^2}{2}\right)}=\frac{\left(\tilde\lambda -\Phi\right)^{3}}{\left(\tilde\lambda +\Phi\right)} \label{Op} \ ,
\een
where we have used the shorthand notations
\begin{eqnarray}
\tilde\lambda&\equiv & \l+\e\k^2V \\
\Phi &\equiv & \tfrac{\e\k^2\dot{\phi} ^2}{2} \ .
\end{eqnarray}
Therefore, we are ready to cast the field equations (\ref{eq:Rmunuq}) for this problem as
\be \label{eq:RmunuOm}
\e {R^\m}_\n (q)=
\(\ba{cc} 1-\frac{\left(\tilde\l+\Phi\right)}{|\O|^{\frac12}}& 0 \\
0 & \left(1-\frac{\left(\tilde\l-\Phi\right)}{|\O|^{\frac12}}\right)\delta^i_j \ea\) \ ,
\ee
where the determinant of the deformation matrix reads $|\O|^{\frac12}=\Omega_+^{1/2}\Omega_{-}^{3/2}$. Now, to compute the left-hand side of the field equations (\ref{eq:RmunuOm}) one can write another line element for the auxiliary metric also of the FLRW form, that is
\begin{eqnarray}\label{dsq2}
ds_q^2 &\equiv& q_{\m\n}dx^{\m}dx^{\n} =-dT^2+\tilde{a}^2(T) d\vec{x}^2   \\
&=&-\O_+(t) dt^2+a^2(t)\O_- (t) \,d\vec{x}^2 \nonumber \ ,
\end{eqnarray}
where in the second line we have used the fundamental relation (\ref{eq:qg}) together with the ansatz (\ref{eq:omcos}). These gravitational field equations must be supplemented with the scalar field equations (\ref{eq:phi}) which, in the FLRW background (\ref{ds2}), read
\begin{equation} \label{eq:scalareq}
\ddot{\phi} + 3 \frac{\dot a}{a} \dot{\phi} + V_\phi=0 \ .
\end{equation}
Now, using the fact that in the $q_{\mu\nu}$ geometry (\ref{dsq2}) we have the well known formulas
\begin{eqnarray}
{R^t}_{t} &\equiv& \frac{3}{\tilde{a}}  \frac{d^2\tilde{a}}{dT^2} \\
{R^i}_{i}  &\equiv&   \frac{1}{\tilde{a}} \frac{d^2\tilde{a}}{dT^2} + \frac{2}{\tilde{a}^2} \left(\frac{d\tilde{a}}{dT}\right)^2 \, , \label{eq:Rii}
\end{eqnarray}
with the relations
\begin{eqnarray}
\frac{d\tilde{a}}{dT}&=& \frac{1}{\Omega_+^{\frac{1}{2}}}\frac{d}{dt}\left(a \Omega_-^{\frac{1}{2}} \right) \\
 \frac{d^2\tilde{a}}{dT^2}&=& \frac{1}{\Omega_+^{\frac{1}{2}}}\frac{d}{dt}\left[\frac{1}{\Omega_+^{\frac{1}{2}}}\frac{d}{dt}\left(a \Omega_-^{\frac{1}{2}} \right)\right] \ ,
\end{eqnarray}
it follows that the combination
\be
3 {R^i}_{i} - {R^t}_{t}  \equiv \frac{6}{\tilde{a}^2} \left(\frac{d\tilde{a}}{dT}\right)^2 \ ,
\ee
can be written as
\be
3 {R^i}_{i} - {R^t}_{t}  \equiv \frac{6}{\Omega_-\Omega_+}\left[\frac{1}{a}\frac{d}{dt}\left(a \Omega_-^{\frac{1}{2}} \right)\right]^2 \ ,
\ee
and a little algebra leads to
\begin{eqnarray}
\frac{1}{a}\frac{d}{dt}\left(a \Omega_-^{\frac{1}{2}} \right)&=&\frac{1}{\Omega_-^{3/2}}\left[\frac{\dot a}{a}\left({\tilde\lambda^2+2\Phi^2}\right)+ {\left(\tilde\lambda +\Phi\right)}\tfrac{\Phi V_\phi}{\dot{\phi}}\right] \ , \nonumber
\end{eqnarray}
which allows us to find an expression for $H \equiv \dot a/a$ as
\begin{equation}\label{eq:H}
H = \frac{1}{\left({\tilde\lambda^2+2\Phi^2}\right)}
\left[- {\left(\tilde\lambda +\Phi\right)}\tfrac{\Phi V_\phi}{\dot{\phi}} \pm  \Omega_-^2\sqrt{\frac{\Omega_+}{3\epsilon}\left( |\O|^{\frac12}-\left(\tilde\l-2\Phi\right)\right)}\right] .
\end{equation}
The square of this quantity is the generalized FLRW equation for EiBI gravity coupled to a scalar field with a potential $V(\phi)$. The $\pm$ signs in front of the square root yield expanding (+) or contracting (-) universes and must be chosen on physical grounds. In the limit $\epsilon\to 0$ the above expression yields
\begin{equation} \label{eq:limittoGR}
H=\pm \sqrt{\frac{\kappa^2(\dot{\phi}^2+2V)}{6}} \ ,
\end{equation}
which recovers the FLRW dynamics of GR coupled to a scalar field.

Note that, from Eq.(\ref{eq:Tttij}), for a scalar field the energy density $\rho_\phi$ and pressure $P_\phi$ are related to the field variables by
\begin{eqnarray}
\rho_\phi&\equiv & \frac{\dot{\phi}^2}{2}+V \\
P_\phi&\equiv & \frac{\dot{\phi}^2}{2}-V \ ,
\end{eqnarray}
which allows to write the quantities inside the functions $\Omega_\pm$ of Eqs. (\ref{Om}) and (\ref{Op}) in terms of $\rho_\phi$ and $P_\phi$ as
\begin{eqnarray}
\tilde\lambda +\Phi&= & \lambda +\epsilon \kappa^2\rho_\phi \label{eq:Omrho} \\
\tilde\lambda -\Phi&= & \lambda -\epsilon \kappa^2P_\phi \label{eq:OmP} \ .
\end{eqnarray}

\section{Massless scalar fields} \label{sec:III}

To explicitly solve the gravitational and scalar field equations (\ref{eq:RmunuOm}) and (\ref{eq:phi})  we need to specify a form of the scalar potential $V(\phi)$. For the sake of simplicity, and to make contact with similar settings in black hole scenarios \cite{Afonso:2017aci}, let us consider first the case of a free scalar field, $V=0$, for which the scalar field equation \eqref{eq:scalareq} has a first integral
\begin{equation}
\dot{\phi}=\frac{\dot{\phi}_0}{a^3(t)} \ ,
\end{equation}
 with $\dot{\phi}_0$ an integration constant. This equation is formally the same as in the GR case, though the scale factor implicitly contains the $\epsilon$-corrections via the resolution of Eq.(\ref{eq:H}), smoothly reducing to their GR values in the limit $\epsilon \to 0$, as follows from Eq.(\ref{eq:limittoGR}).
 
 Since in this case $P_\phi=\rho_\phi=\dot{\phi}^2/2$, the components of the deformation matrix (\ref{eq:omcos}) assume the relatively simple form
\begin{equation}
\O_-=\(\l^2-\frac{\rho_\phi^2}{\rho_\epsilon^2} \)^{\frac12} \hspace{0.1cm}; \hspace{0.1cm}
\O_+=\frac{\left(\lambda-s\frac{\rho_\phi}{\rho_\epsilon}\right)^{3/2} }{\left(\lambda+s\frac{\rho_\phi}{\rho_\epsilon}\right)^{1/2}} \ ,
\end{equation}
where we have defined the critical density $\rho_\epsilon=1/(\kappa^2|\epsilon|)$ and $s=\pm 1$ denotes the sign of $\epsilon$. The corresponding Hubble function takes the form
\begin{equation}
H^2 = \frac{\left(\lambda ^2-\frac{\rho^2_\phi}{\rho^2_\epsilon}\right)^{3/2} \left(\lambda -s \frac{\rho_\phi}{\rho_\epsilon}\right)^2}{3 s \epsilon  \left(\lambda ^2+2 \frac{\rho^2_\phi}{\rho^2_\epsilon}\right)^2} \label{eq:Hubfact}  \left(  \left(\lambda -s\frac{\rho_\phi}{\rho_\epsilon}\right)\sqrt{\lambda ^2-\frac{\rho^2_\phi}{\rho^2_\epsilon}}-\lambda+2s \frac{\rho_\phi}{\rho_\epsilon}\right) .
\end{equation}

As mentioned before, at low densities as compared to the scale $\rho_\epsilon$, this equation recovers the GR expression, which can be solved leading to the approximate solution $a(t)=H_0 t^{1/3}$. The Hubble factor here can be conveniently set to $H_0^2 = 3\kappa^2 \dot{\phi}^2_0/2$  to adjust the integration constant $\dot{\phi}_0$ in order to reproduce the standard cosmological evolution of GR coupled to a massless scalar field. At higher densities, however, the dynamics strongly departs from that of GR and it is evident that when the energy density of the scalar field approaches its maximum value, $\rho_\phi=\lambda \rho_\epsilon$, the Hubble function vanishes. This corresponds to a minimum value of the expansion factor for both branches of solutions $s=\pm 1$, but there are two different mechanisms by which the initial Big Bang singularity is avoided, which we discuss next. For simplicity, from now on we focus on asymptotically flat configurations, $\lambda=1$.

\begin{figure}[t!]
\centering
\includegraphics[width=8.0cm,height=6.0cm]{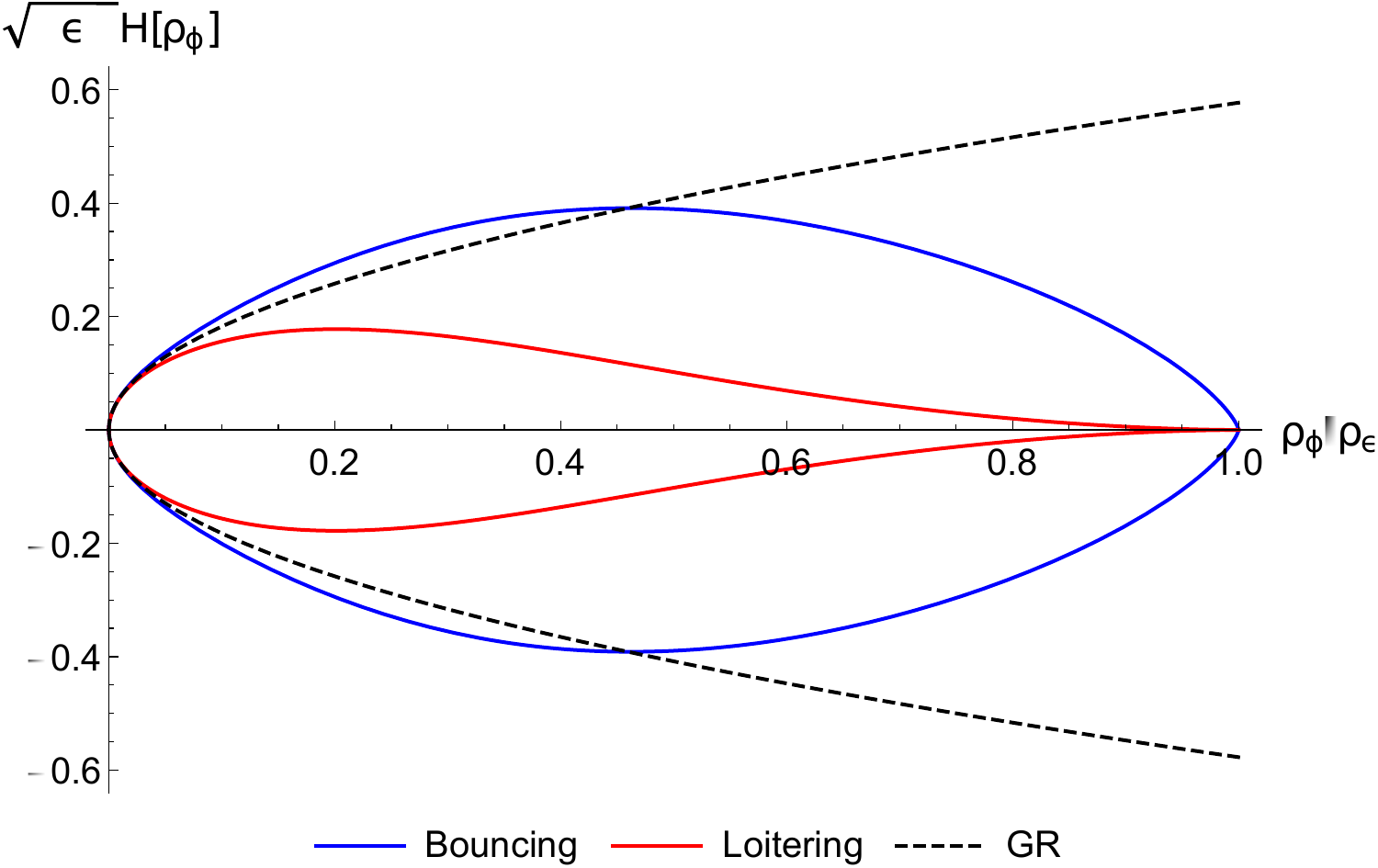}
\caption{Representation of the Hubble function  $\sqrt{ \vert \epsilon \vert }H(\rho_\phi)$ vs. the energy density for EiBI gravity (solid) and GR (dashed) with massless scalar field as a function of the ratio $\rho_\phi/\rho_\epsilon$. Note how the trajectories of the EiBI theory are bounded, while that of GR (dashed black) is open. Bouncing solutions (blue) reach the maximum density forming a $\pi/2$ angle with the horizontal axis, while the loitering branch (red) reaches it tangentially.}
\label{fig:PS}
\end{figure}

For the $s=-1$ branch, at the critical density $\rho_{\phi}=\rho_{\epsilon}$,  the Hubble factor in Eq.(\ref{eq:Hubfact}) and its derivative behave as
\begin{eqnarray}
H(\rho_{\phi})&\approx & \pm \frac{2^{7/4}}{3|\epsilon|^{1/2}}\left(1-\frac{\rho_\phi}{\rho_\epsilon}\right)^{3/4} \\
\frac{dH}{d\rho_\phi} &\approx &\mp \frac{1}{2^{\frac{1}{4}}|\epsilon|^{\frac{1}{2}}\rho_\epsilon\left(1-\frac{\rho_\phi}{\rho_\epsilon}\right)^{1/4}} \ .\end{eqnarray}
This indicates that at the maximum achievable density, $\rho_\phi=\rho_\epsilon$, the Hubble factor vanishes while its derivative goes to infinity. This implies that the universe contracts to a minimum (maximum) value of the expansion factor (energy density), before re-expanding. This the typical behaviour expected in standard bouncing solutions with a transition from a contracting phase to an expanding one, where the universe could even undergo a sequence of cyclic cosmologies with both phases. In Fig. \ref{fig:PS} we plot the form of $H(\rho)$ (blue curves) according to Eq.(\ref{eq:Hubfact}) to show that while at late-times (i.e. low densities, $\rho_{\phi} \ll \rho_{\epsilon}$) these solutions converge to those of GR (in agreement to Eq.(\ref{eq:limittoGR})),  at high densities ($\rho_{\phi}/\rho_{\epsilon} \to 1$) they depart form the standard Big Bang singularity of GR.

For the $s=+1$ branch we find that the evolution of the Hubble factor as we approach the maximum density $\rho_\phi=\rho_\epsilon$ is given instead by
\begin{eqnarray}
H(\rho_\phi) &\approx & \pm \frac{2^{3/4} \sqrt{\frac{1}{\epsilon }} \left(1-\frac{\rho_\phi }{\rho_ \epsilon }\right)^{7/4}}{3 \sqrt{3}} \\
\frac{dH}{d\rho_\phi} &\approx&  \mp \frac{7 \sqrt{\frac{1}{\epsilon }} \left(1-\frac{\rho_\phi }{\rho_\epsilon }\right)^{3/4}}{6 \sqrt[4]{2} \sqrt{3} \rho_\epsilon } \ ,
\end{eqnarray}
which means that the Hubble factor vanishes there as well, but instead of being divergent its derivative takes a finite value (which is actually zero). This implies that the expansion factor reaches a fixed value $a(t)=a_m$, corresponding to an asymptotically Minkowski past, and starts expanding as we move forward in time after some reference time $t=t_p$. In Fig. \ref{fig:PS}  we depict this behaviour (red curves) of these so-called loitering solutions, where the qualitative differences with the bouncing solutions are manifest. Again, at late times (low densities) the standard GR evolution is recovered. Therefore we see that a free massless scalar field coupled to EiBI gravity is able to yield nonsingular evolutions in both $s=\pm 1$ branches according to these two mechanisms, something not possible within GR.

\section{Massive scalar field}  \label{sec:IV}

Let us now consider a massive scalar field with a potential of the form
\begin{equation}
V(\phi)=\frac{1}{2}\mu^2\phi^2 \ ,
\end{equation}
where $\mu$ is a constant. This choice for the potential, besides being the simplest form, corresponds to the natural mass associated to a canonical scalar field. In this case, the Hubble function (\ref{eq:H}) develops an explicit dependence on both $\rho_\phi$ and $P_\phi$ [to lighten the notation from now on we shall drop the label $\phi$ in these quantities] via Eqs.(\ref{eq:Omrho}) and (\ref{eq:OmP}). Since the resulting expression does not provide any useful insight, we will not write it explicitly here. Instead, for the sake of comparison with the massless case of the previous section, we find it more illustrative to provide a parametric representation of $H(t)$ versus $\rho(t)$. These functions are obtained by numerically integrating the second-order equation for $\ddot{a}$ that follows from (\ref{eq:Rii}) together with its corresponding right-hand side. The result of the integration is used to construct the quantity $\dot{a}/a$, which is then compared with the formula (\ref{eq:H}) to check the consistency of the numerical integration. Again we split our discussion of the corresponding results into the $s=\pm 1$ cases.

\begin{figure}[t!]
\centering
\includegraphics[width=0.6\textwidth]{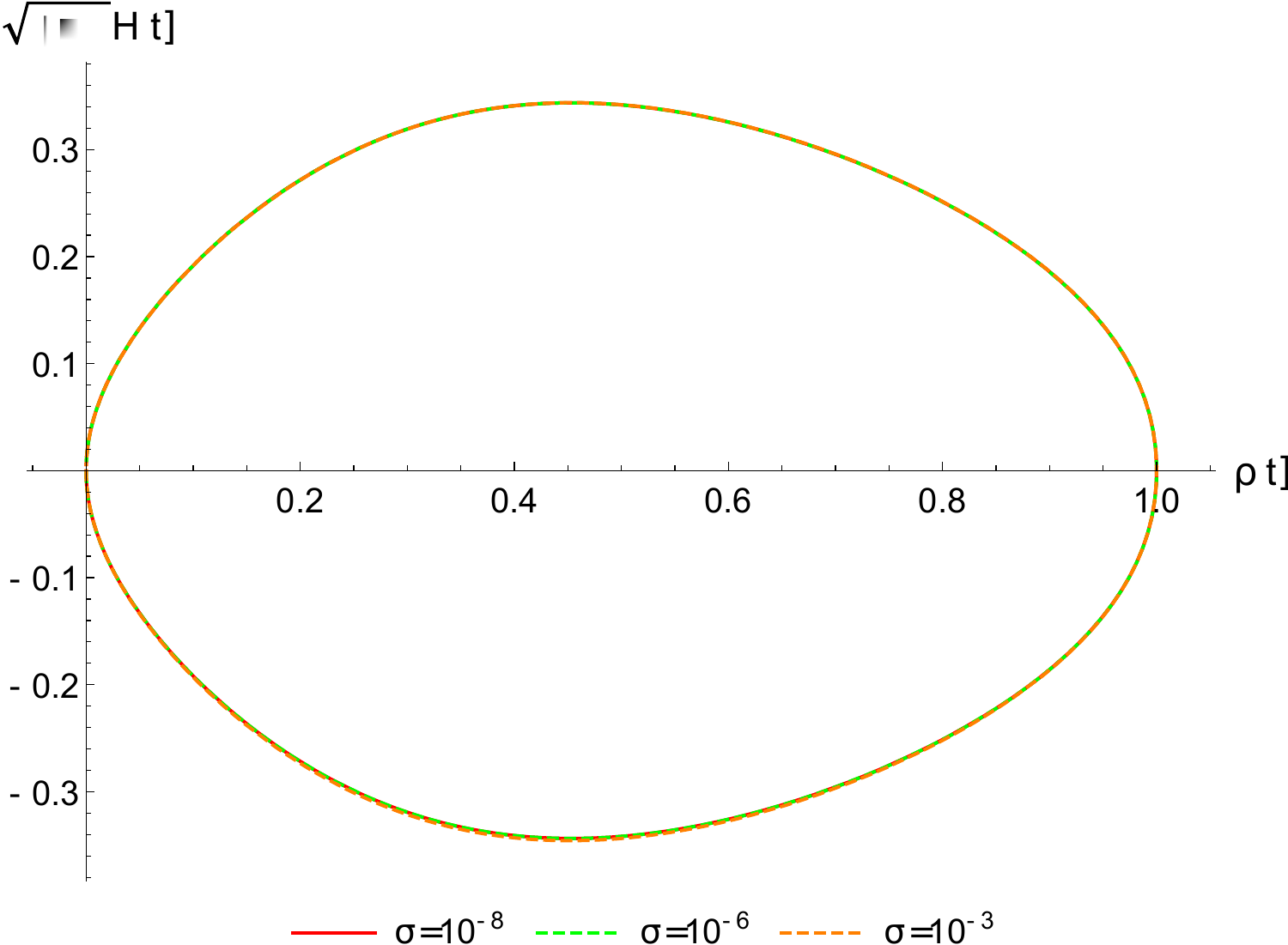}
\caption{Parametric plot of $\sqrt{|\epsilon|}H(t)$ as a function of $\rho(t)/\rho_\epsilon$ for small masses when $s=-1$ (bouncing solutions). The (reduced) mass parameter is taken such that $\sigma\equiv \mu^2/\rho_{\epsilon}$. }
\label{fig:BHrhoS}
\end{figure}

\begin{figure}[t!]
\centering
\includegraphics[width=0.6\textwidth]{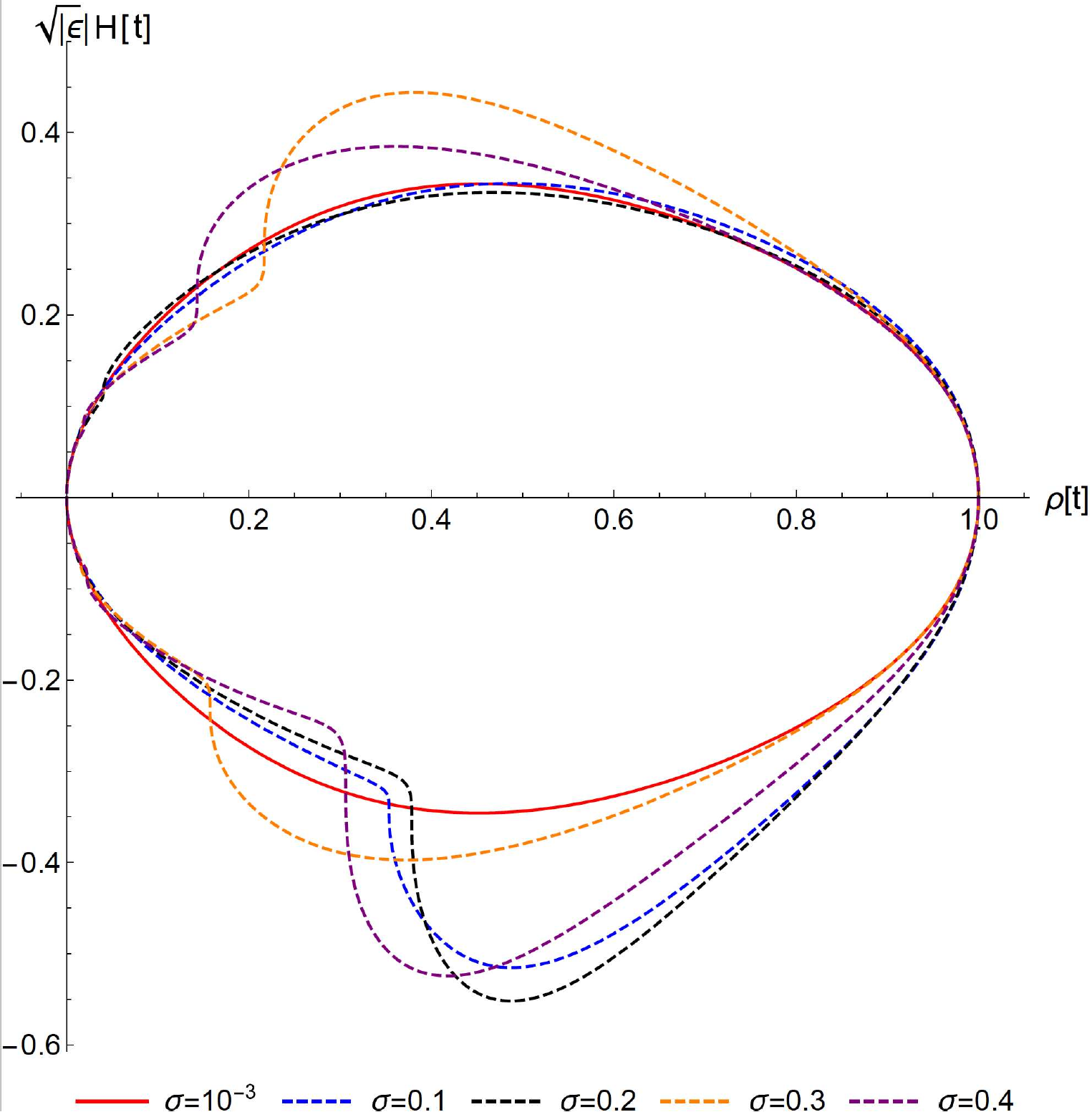}
\caption{Parametric plot of $\sqrt{|\epsilon|}H(t)$  as a function of $\rho(t)$ for larger values of the reduced mass $\sigma\equiv \mu^2/\rho_{\epsilon}$ when $s=-1$. As compared to Fig. \ref{fig:BHrhoS}, in this case the development of fish-like structures is clearly visible as $\sigma$ grows large enough.}
\label{fig:BHrhoL}
\end{figure}

\begin{figure}[t!]
\centering
\includegraphics[width=0.6\textwidth]{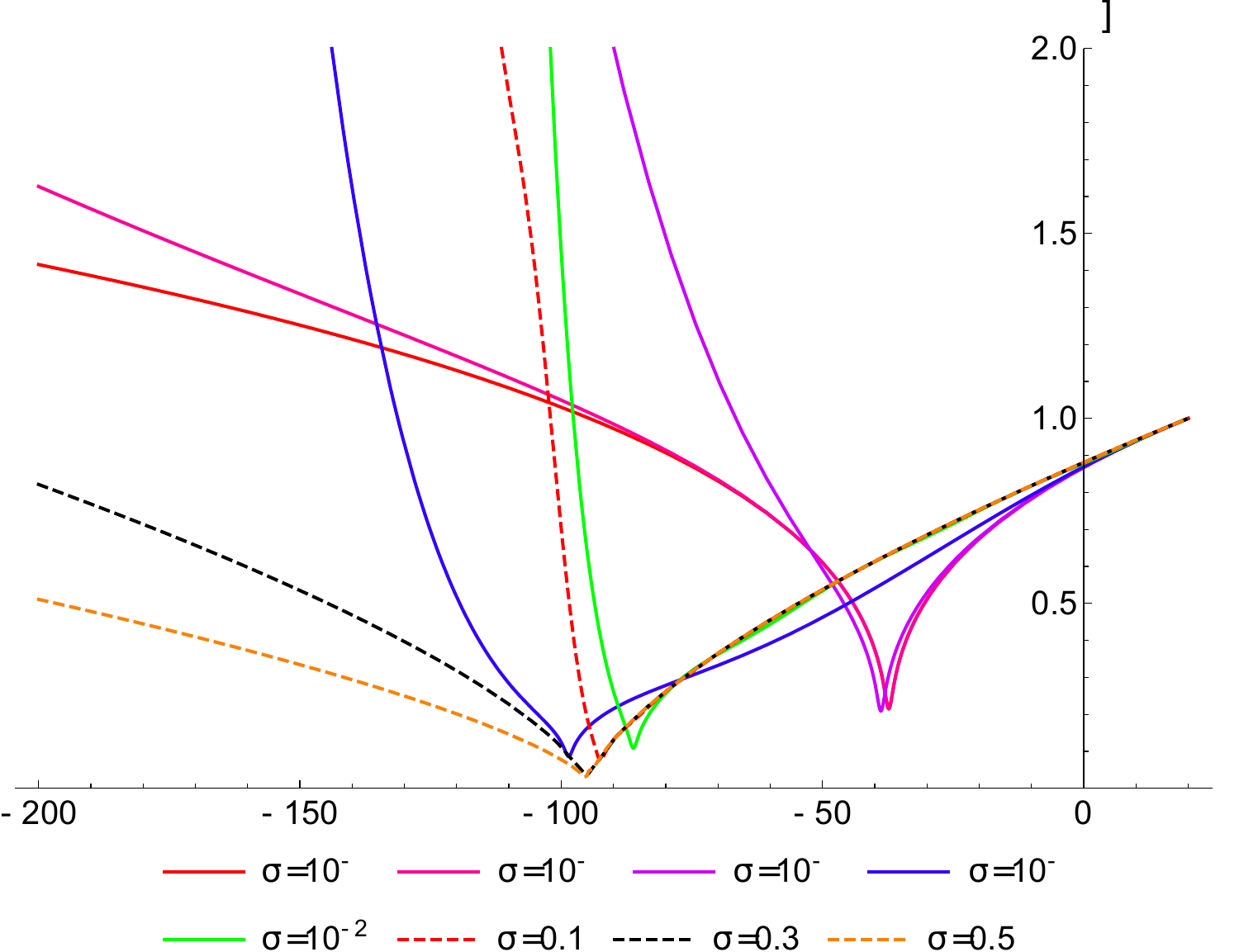}
\caption{Expansion factor $a(t)$ representing bouncing solutions for various values of the reduced mass parameter $\sigma\equiv \mu^2/\rho_{\epsilon}$ when $s=-1$.}
\label{fig:Ba}
\end{figure}

\begin{figure}[h!]
\centering
\includegraphics[width=0.6\textwidth]{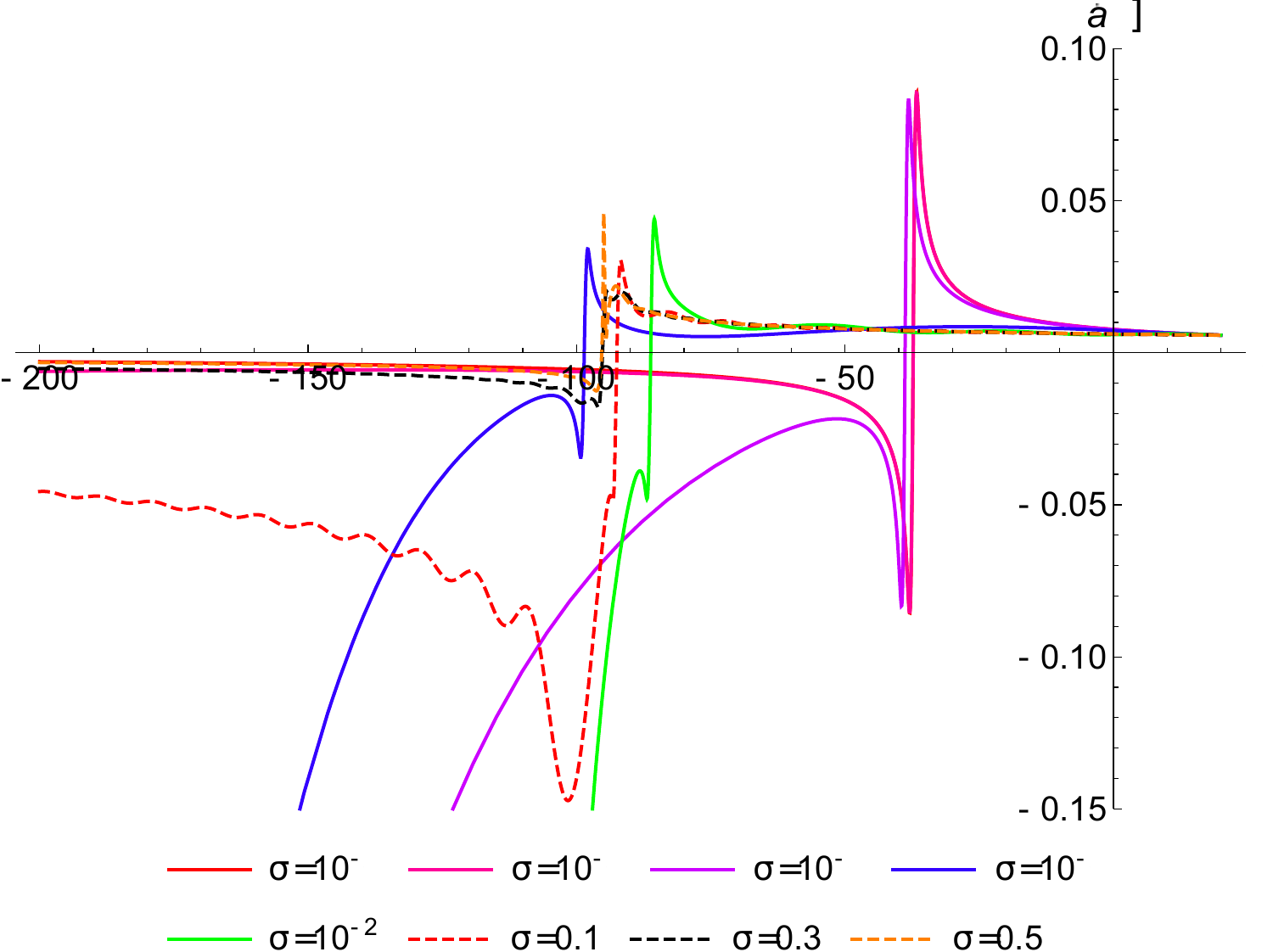}
\caption{Time derivative of the expansion factor representing bouncing solutions for various values of the reduced mass parameter $\sigma\equiv \mu^2/\rho_{\epsilon}$. The colors of the curves are the same as in Fig. \ref{fig:Ba}. Note that the blue and magenta curves will bounce at some point and begin a growing oscillatory trajectory similar to that of the red dashed curve, being this a generic behavior of all these solutions.}
\label{fig:Bap}
\end{figure}

\begin{figure}[h!]
\centering
\includegraphics[width=0.6\textwidth]{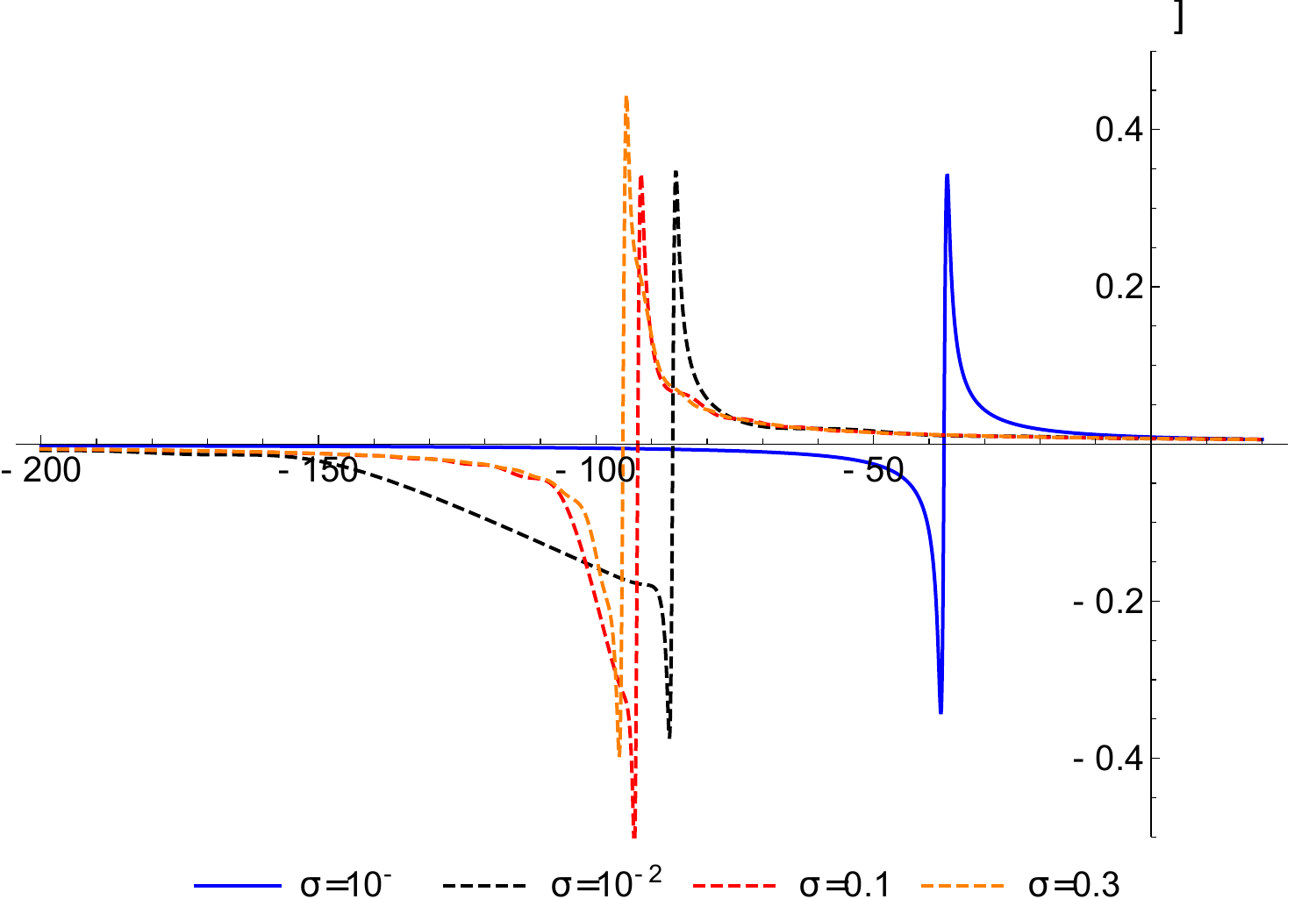}
\caption{Hubble function $H \equiv \dot{a}/a$ for various values of the reduced mass parameter $\sigma\equiv \mu^2/\rho_{\epsilon}$ of the $s=-1$ case.}
\label{fig:BH}
\end{figure}

\begin{figure}[h!]
\centering
\includegraphics[width=0.6\textwidth]{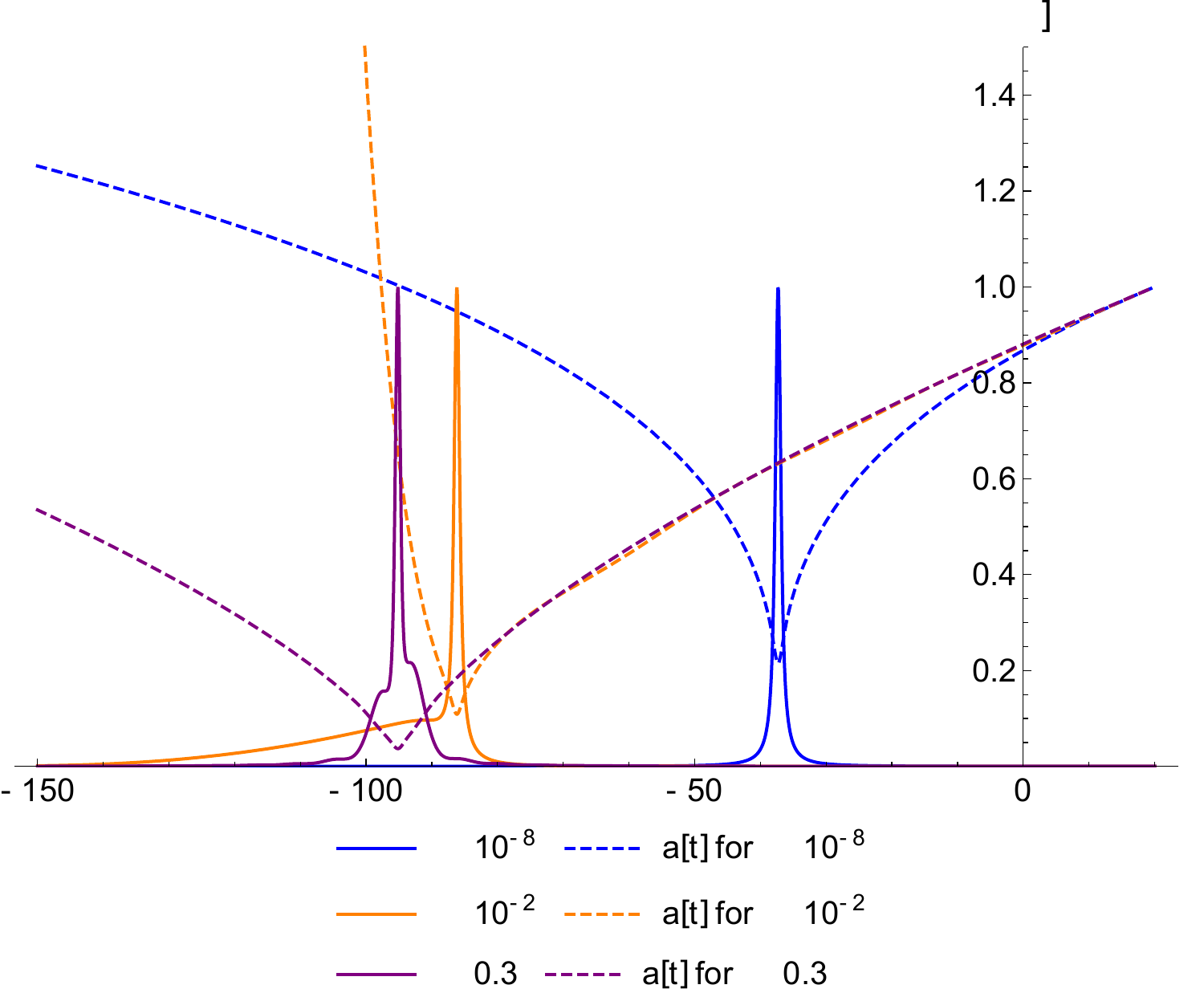}
\caption{Energy density (solid curves) as a function of time superimposed with its corresponding expansion factor $a(t)$ (dashed curves) for some bouncing solutions of the $s=-1$ case.}
\label{fig:Brho}
\end{figure}

 In Figs. \ref{fig:BHrhoS} and \ref{fig:BHrhoL}, we show the function $H(\rho(t))$ from Eq.(\ref{eq:H}) with the help of (\ref{Om}) and (\ref{Op}) particularized to the potential above, for solutions corresponding to several values of the reduced mass $\sigma \equiv \mu^2/\rho_{\epsilon}$ in the case $s=-1$. The corresponding expansion factors $a(t)$ [obtained from integrating the previous function $H$] appear in Fig. \ref{fig:Ba}, while in Fig. \ref{fig:Bap} we show the function $\dot{a}(t)$ of those same examples and in Fig. \ref{fig:BH} their Hubble functions. As one can see from all these plots, the case $s=-1$ still represents bouncing solutions  for all the mass range explored, which goes from  $\sigma \approx 10^{-8}$ up to $\sigma\approx 0.4$, with little variations with respect to the massless scenario for masses as high as $\sigma\approx 10^{-3}$ (see Fig. \ref{fig:BHrhoS}). For higher masses, the egg-shaped Hubble function develops a fish-like structure, with asymmetric fins. This asymmetry is also manifest in the contracting branch of the expansion factor (see Fig. \ref{fig:Ba}), which becomes increasingly asymmetric as the mass parameter grows from zero. A similar behaviour can be observed in the density profiles of the scalar field depicted in Fig. \ref{fig:Brho}.

\begin{figure}[h!]
\centering
\includegraphics[width=0.6\textwidth]{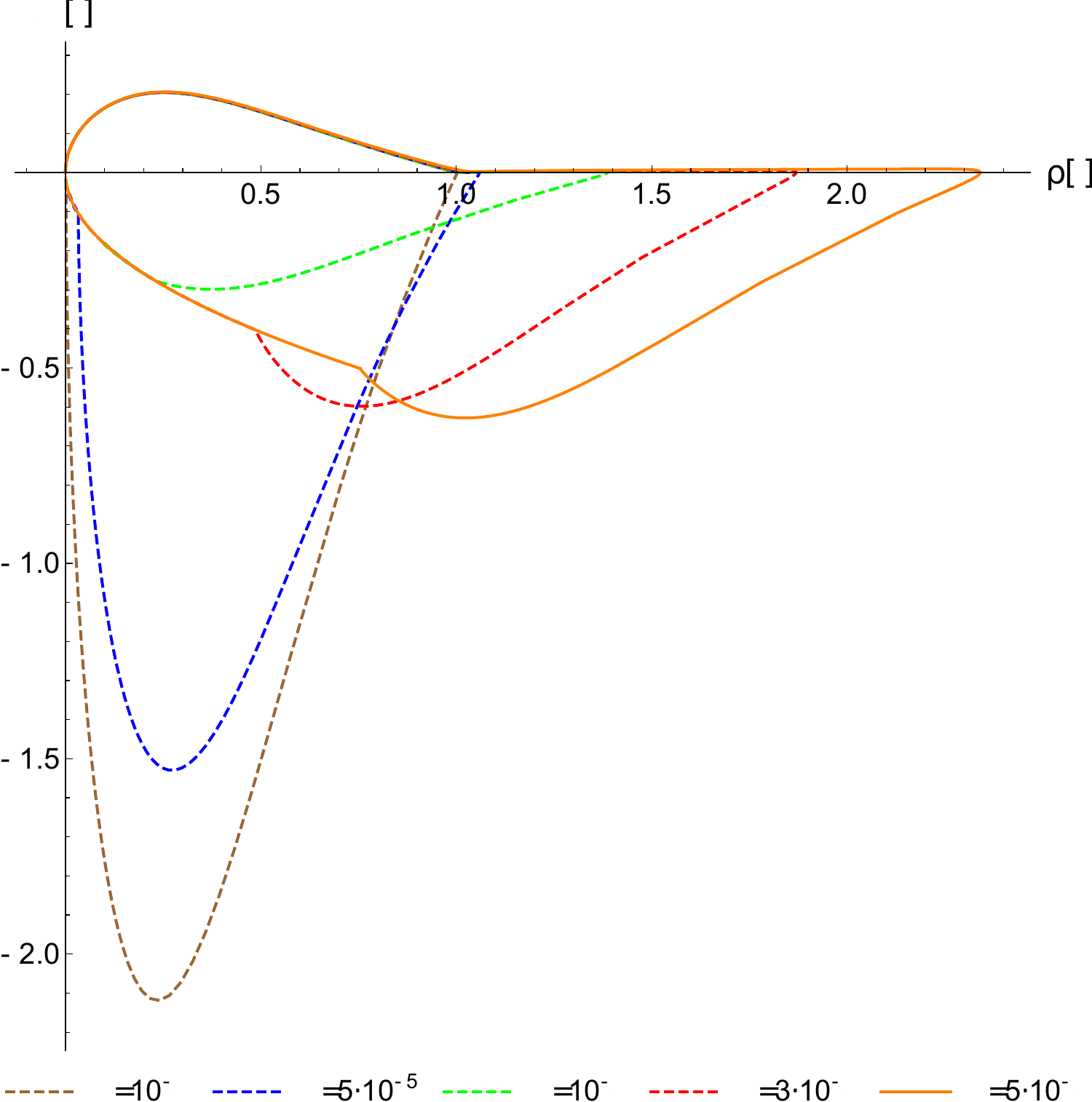}
\caption{Parametric plot of $H(t)$ as a function of $\rho(t)$ for small values of the reduced mass $\sigma$ of the $s=+1$ case.}
\label{fig:BHrhoSL}
\end{figure}

\begin{figure}[h!]
\centering
\includegraphics[width=0.6\textwidth]{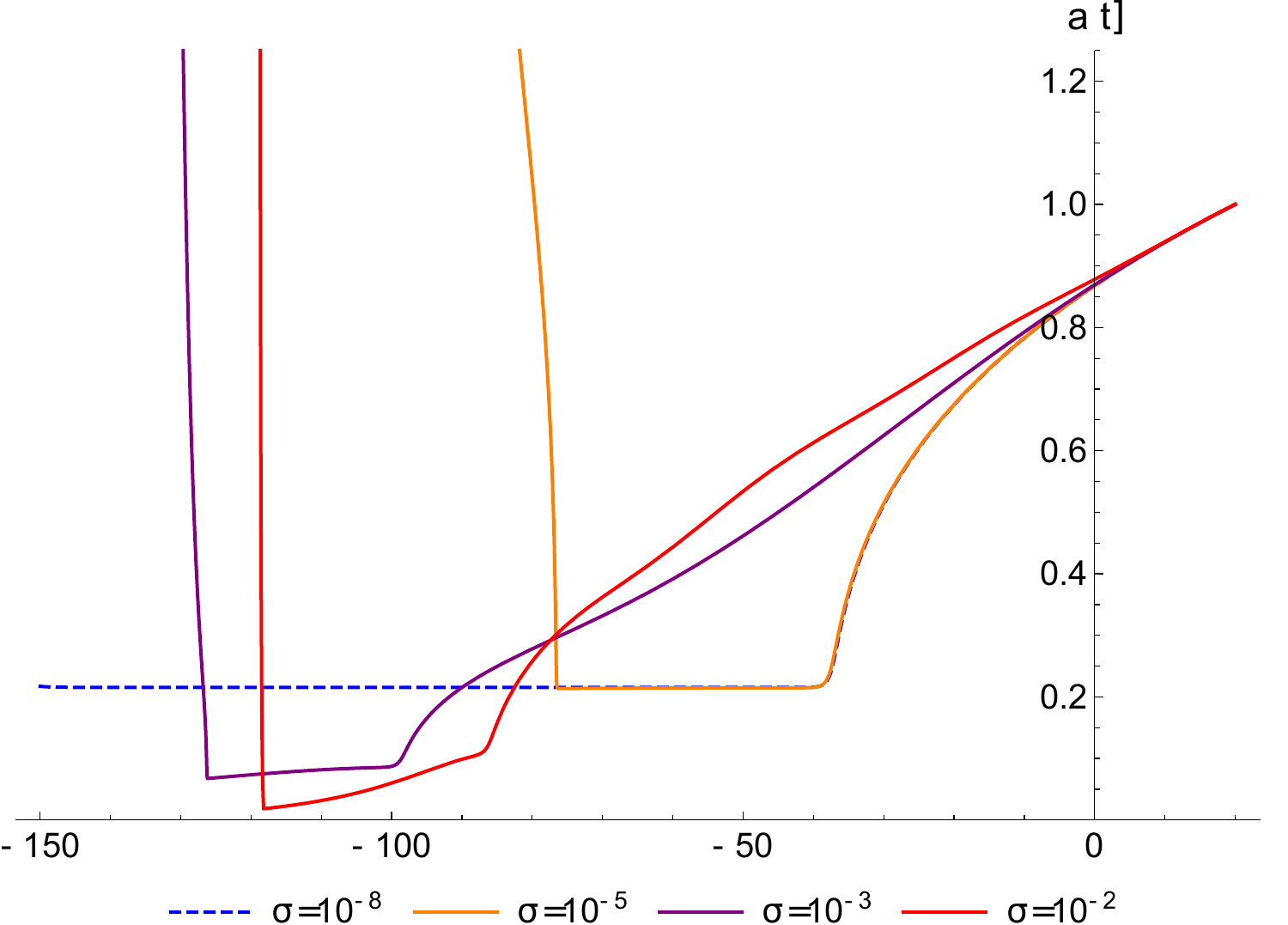}
\caption{Expansion factor representing new bouncing solutions for various values of the reduced mass parameter $\sigma$ when $s=+1$. }
\label{fig:BaL}
\end{figure}

\begin{figure}[h!]
\centering
\includegraphics[width=0.6\textwidth]{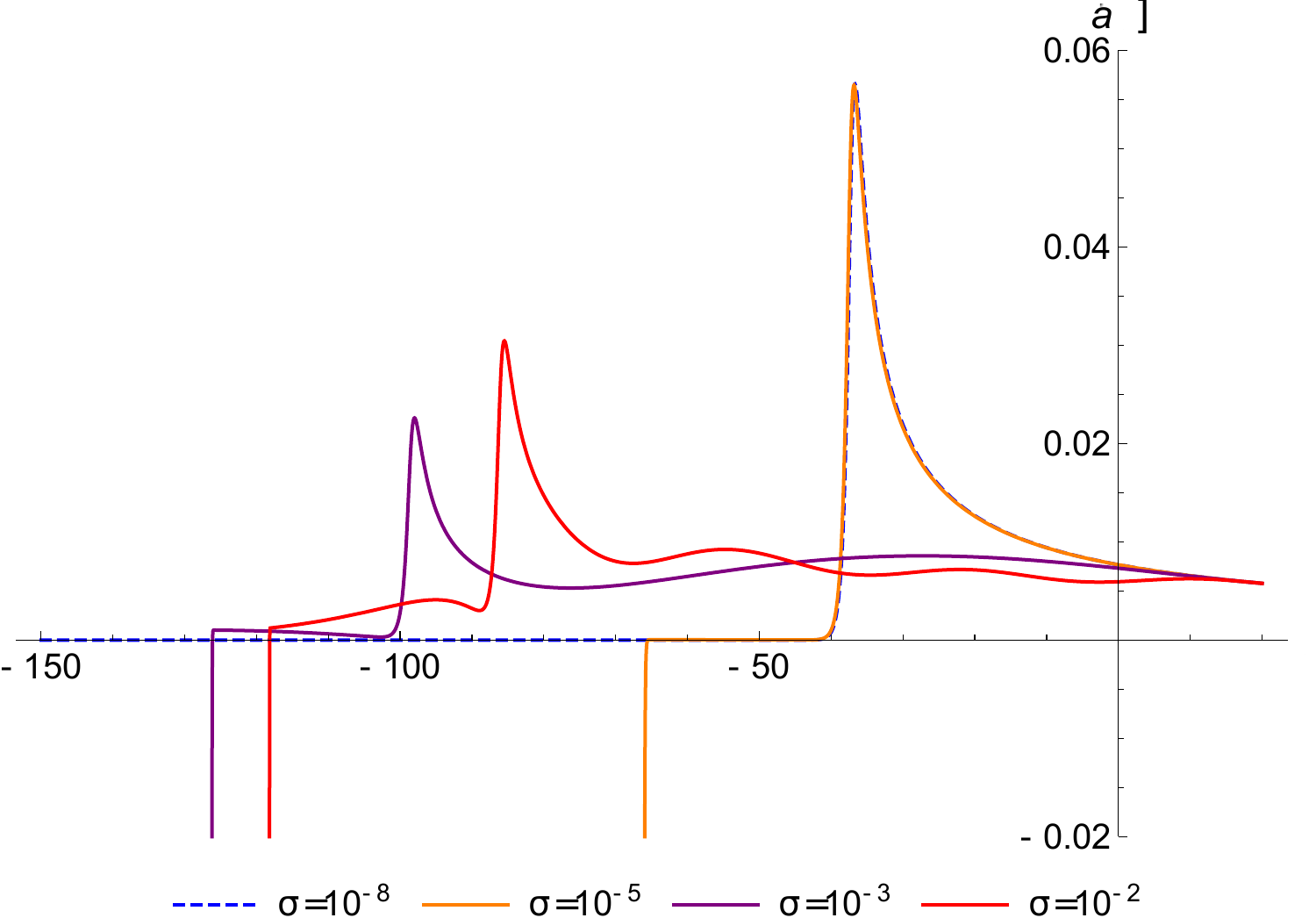}
\caption{Time derivative of the expansion factor representing bouncing solutions for the reduced mass parameter $\sigma$ when $s=+1$.}
\label{fig:BapL}
\end{figure}

\begin{figure}[h!]
\centering
\includegraphics[width=0.6\textwidth]{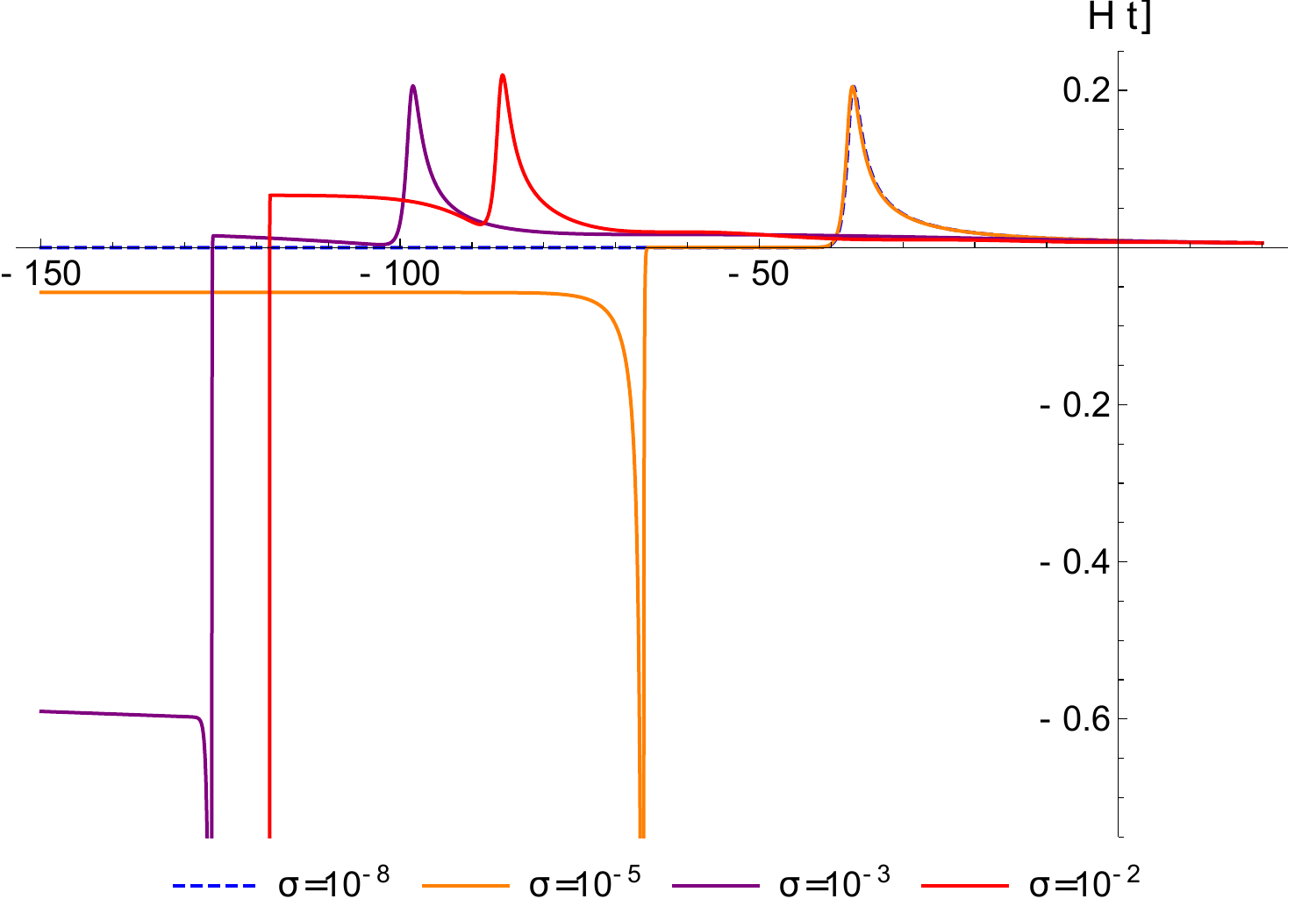}
\caption{Hubble function $\dot{a}/a$ for various values of the reduced mass parameter $\sigma$ when $s=+1$.}
\label{fig:BHL}
\end{figure}

\begin{figure}[h!]
\centering
\includegraphics[width=0.6\textwidth]{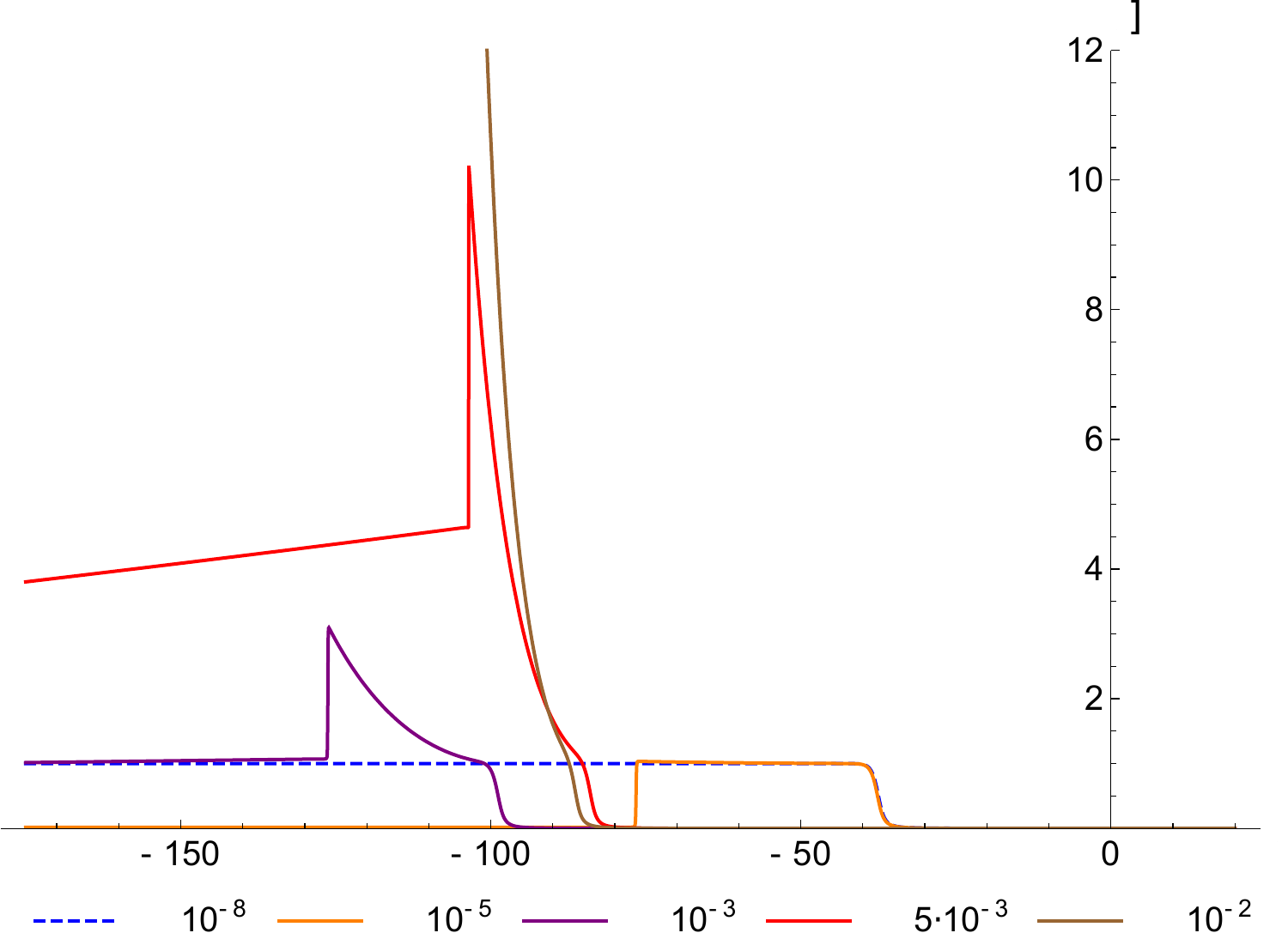}
\caption{Energy density (solid curves) as a function of time superimposed with its corresponding expansion factor $a(t)$ (dashed curves).}
\label{fig:BrhoLI}
\end{figure}

\begin{figure}[h!]
\centering
\includegraphics[width=0.6\textwidth]{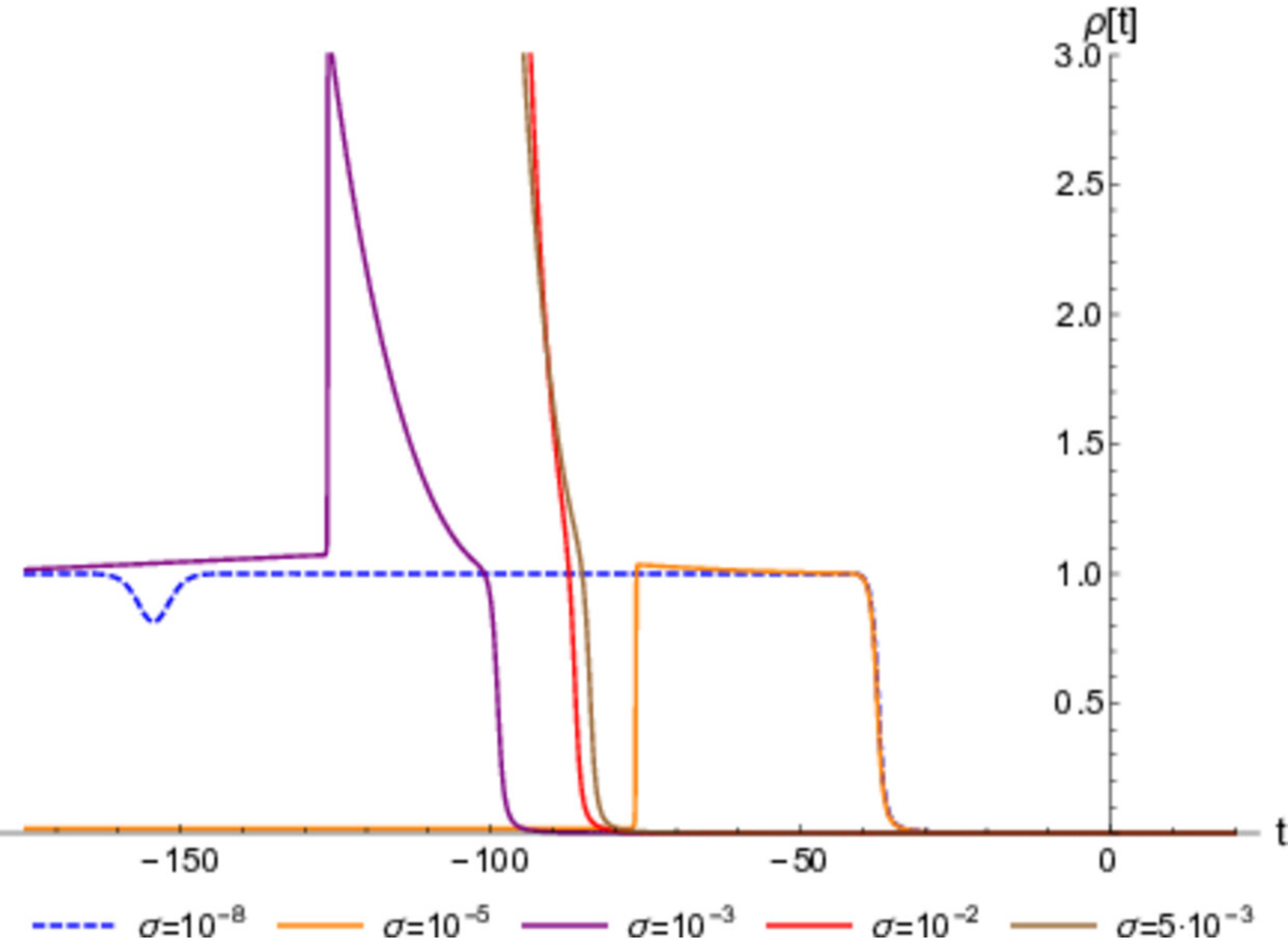}
\caption{Zoom in of the energy density (solid curves) as a function of time superimposed with its corresponding expansion factor $a(t)$ (dashed curves).}
\label{fig:BrhoLII}
\end{figure}

Focusing now on the case with $s=+1$, which represents loitering solutions in the massless case,  Fig. \ref{fig:BHrhoSL} shows that the function $H(\rho(t))$ is now qualitatively different (compare to Fig. \ref{fig:PS}). Indeed, when $\sigma=0$, $H(t)$ never crosses the horizontal axis as one approaches the maximum allowed energy density. However, for $\sigma \neq 0$ one either has an expanding branch which starts from an almost constant $a(t)=a_m$ initial phase or a contracting phase that ends up in an almost constant final $a(t)=a_m$. Now we observe that {\it all} solutions develop a bounce at some high density, effectively crossing the axis and establishing a continuous transition from a contracting phase to an expanding one. The loitering phase, with an almost constant $a(t)$, may last for a long period after an initial contraction but it will always end up in an expanding branch. This is illustrated in Fig. \ref{fig:BaL}, where  the blue dashed curve represents a phase with $a(t)$ almost constant for a long time (very low mass). The orange curve has essentially the same future behaviour as the blue curve but its loitering phase (almost constant expansion factor) does not last so much, exhibiting a previously contracting phase that arises abruptly. Similar behaviours arise for larger masses, and one  verifies that the instabilities in the expansion factor always lead to a bounce, never finding fully collapsed solutions. This is a remarkable property of the EiBI model, since it always avoids the development of singular solutions.

For completeness, the behaviour of $\dot{a}(t)$ and $H(t)$ for this case $s=+1$ are shown in Figs. \ref{fig:BapL} and \ref{fig:BHL}. The corresponding energy density profiles appear in Figs. \ref{fig:BrhoLI} and \ref{fig:BrhoLII}. It is amusing to see how for low mass configurations (orange solid curve, for instance) the energy density at early times is very low (large universe with very diluted energy) until it rapidly increases to reach a maximum, where it stays for some time at an almost constant value (loitering phase) with a slight decay (decompression)  as we move forward in time. Then the density suddenly drops again giving rise to an expanding phase. For larger values of the scalar field mass, this process can be significantly deformed but the qualitative features remain.

\section{Observational Constraints}
\label{sec:obs}

In this section we shall put to observational test the EiBI cosmologies considered in the previous sections by using the latest observational data to constrain the EiBI parameter $\epsilon$ preventing the singularity in the early Universe.
\subsection{BBN}

In the first sections we began with the original action with the scalar field inside, and shifted the solution to other frame. This frame, as we showed, removes the singularity of the solution and for $\epsilon \rightarrow 0$ it recovers GR. Here, we put the density of the $\Lambda$CDM that produces in the "Einstein frame" a solution that still removes the singularity and recovers GR for $\epsilon \rightarrow 0$. Doing the same procedure for a theory with matter fields $\rho_m$ in the original frame gives the density:
\begin{equation}
\tilde{\rho} = \frac{8 (3 p \epsilon -1)^2 (3 \rho  \epsilon +1) \left(2 \sqrt{1-3 p \epsilon } \sqrt{3 \rho  \epsilon +1}-3 \epsilon  \left(p \left(2 \sqrt{1-3 p \epsilon } \sqrt{3 \rho  \epsilon +1}-3\right)-\rho \right)-2\right)}{3 \epsilon  \left((3 \rho  \epsilon +1) \left(3 \epsilon  \left(a p'+4 p\right)-4\right)+3 a \epsilon  (3 p \epsilon -1) \rho' \right)^2}
\end{equation}
in the physical frame. For $\epsilon \rightarrow 0$ the relation gives $\tilde{\rho} = \rho$, as expected. 

In order to constrain the EiBI parameter $\epsilon$ we use different measurements from our Universe. The strongest one is from the Big Bang Nucleosynthesis (BBN) where $z \sim 10^9$. In principle the condition for the BBN constraint is \cite{Sola:2016jky,Barrow:2020kug}:
\begin{equation}
\left| \frac{\rho_{EiBI} - \rho_{\Lambda CDM} }{\rho_{\Lambda CDM}} \right| < 10 \% 
\label{eq:BBNcon}
\end{equation}
In this era the matter dominance is radiation, so we take the density and the pressure to be:
\begin{equation}
\rho = \frac{\Omega_r}{a^4}, \quad p = \frac{\rho}{3}.
\end{equation}
The density for the EiBI theory gives:
\begin{equation}
\tilde{\Omega} = \frac{\left(a^4-\Omega_r \epsilon \right)^2 \left(a^4+3 \Omega_r \epsilon \right) \left(a^4 \left(\sqrt{1-\frac{\Omega_r \epsilon }{a^4}} \sqrt{\frac{3 \Omega_r \epsilon }{a^4}+1}-1\right)+\Omega_r \epsilon  \left(3-\sqrt{1-\frac{\Omega_r \epsilon }{a^4}} \sqrt{\frac{3 \Omega_r \epsilon }{a^4}+1}\right)\right)}{3 \epsilon  \left(a^8+3 \Omega_r^2 \epsilon ^2\right)^2}.
\end{equation}
From the condition Eq.(\ref{eq:BBNcon}) we get the bound\footnote{Recall that restoring dimensions we have $\epsilon\to \kappa^2\epsilon$, which represents the inverse of a matter density. Accordingly, the bound on $\epsilon$, which represents a squared length, can be written also as $\left|\epsilon\right| \lesssim  5.53 \cdot 10^{-8} \text{ m}^2$. }:
\begin{equation}
\left|\kappa^2  \epsilon\right| \lesssim  1.03 \cdot 10^{-33}.
\end{equation}
where we took $\Omega_r \sim 10^{-4}$. It is important to note that this bound improves any previous existing bounds on $\epsilon$ by several orders of magnitude. In fact, the most stringent constraints so far came from elementary particles scattering experiments \cite{Jimenez:2021sqd,Latorre:2017uve,Delhom:2019wir} and implied $\kappa^2 \epsilon\lesssim 1.86 \cdot 10^{-28}$ (or equivalently, $\epsilon\lesssim 10^{-2} \text{ m}^2$).\\

We now proceed to describe the observational data sets along with the relevant statistics in constraining the model, while we use the first constraint from the BBN above. The matter fields considered are the dark energy $\Omega_\Lambda$, dark matter $\Omega_m$, and radiation $\Omega_r$ components. The corresponding energy density reads
\begin{equation}
\frac{\rho}{\rho_c} = \frac{\Omega_m}{a^3} + \frac{\Omega_r}{a^4} + \Omega_\Lambda \ .
\end{equation}
Inserting this equation in the expression of the Hubble factor (\ref{eq:H}), and assuming a small EiBI parameter, $\vert R_{\mu\nu} \vert \ll \epsilon^{-1}$, yields the extended Friedman equation:
\begin{equation}
 \frac{H^2}{H_0^2} = \frac{\Omega_m}{a^3} + \frac{\Omega_r}{a^4} + \Omega_\Lambda + \epsilon \left( \frac{9 \Omega_m^2}{8 a^6}+\frac{\Omega_m \Omega_r}{a^7}-\frac{2 \Omega_\Lambda  \Omega_r}{a^4} \right) + \mathcal{O} (\epsilon^2) ,
\end{equation}
where the EiBI corrections to GR solutions are apparent.

\subsection{Direct measurements of the Hubble expansion}

Cosmic Chronometers (CC): This data set exploits the evolution of differential ages of passive galaxies at different redshifts to directly constrain the Hubble parameter \cite{Jimenez:2001gg}. We use uncorrelated 30 CC measurements of $H(z)$ discussed in \cite{Moresco:2012by,Moresco:2012jh,Moresco:2015cya,Moresco:2016mzx}. The corresponding $\chi^2_{H}$ function reads:
\begin{equation}
\chi^{2}_{H} = \sum_{i=1}^{30}\left(\frac{H_{i} - H_{pred}(z_i)}{\Delta H_i}\right)^2,
\end{equation}
where $H_{i}$ is the observed Hubble rates at redshift $z_{i}$ ($i=1,...,N$) and $H_{pred}$ is the predicted one from the model.

\subsection{Standard Candles}

As Standard Candles (SC) we use measurements of the Pantheon type Ia supernova \cite{Scolnic:2017caz}, that were collected in \cite{Anagnostopoulos:2020ctz}, as well as quasars \cite{Roberts:2017nkm} and gamma ray bursts \cite{Demianski:2016zxi}. The model parameters are fitted by comparing the observed $\mu _{i}^{obs}$ value to the theoretical $\mu _{i}^{th}$ value of the distance moduli, which are the logarithms:
\begin{equation}
 \mu=m-M=5\log _{10}(D_{L})+\mu _{0} \ ,
\end{equation}
where $m$ and $M$ are the apparent and absolute magnitudes and $\mu_{0}=5\log \left( H_{0}^{-1}/Mpc\right) +25$ is the nuisance parameter that
has been marginalized. The luminosity distance is defined by
\begin{eqnarray}
D_L(z) &=&\frac{c}{H_{0}}(1+z)\int_{0}^{z}\frac{dz'}{%
E(z')} \ ,
\end{eqnarray}%
(here $\Omega _{k}=0$, i.e., a flat space-time and $E(z)$ is the dimensionless Hubble parameter). For the SnIa data the covariance matrix is not diagonal and the distance modulus is given as $\mu_{i} = \mu_{B,i}-\mathcal{M}$, where $\mu_{B,i}$ is the apparent magnitude at maximum in the rest frame for redshift $z_{i}$ and $\mathcal{M}$ is treated as a universal free parameter, quantifying various observational uncertainties \cite{Scolnic:2017caz}. Following standard lines, the chi-square function of the standard candles is given by
\begin{equation}
\chi^{2}_{\text{SC}}\left(\phi^{\nu}_{\text{s}}\right)={\mu}_{\text{s}}\,{C}_{\text{s},\text{cov}}^{-1}\,{\mu}_{\text{s}}^{T}\,,
\end{equation}
where ${\mu}_{\text{s}}=\{\mu_{1}-\mu_{\text{th}}(z_{1},\phi^{\nu})\,,\,...\,,\,\mu_{N}-\mu_{\text{th}}(z_{N},\phi^{\nu})\}$ and the subscript $\text{`s'}$ denotes SnIa and QSOs.

\subsection{Baryon acoustic oscillations}

We use uncorrelated data points from different Baryon Acoustic Oscillations (BAO). BAO are a direct consequence of the strong coupling between photons and baryons in the pre-recombination epoch. After the decoupling of photons, the over densities in the baryon fluid evolved and attracted more matter, leaving an imprint in the two-point correlation function of matter fluctuations with a characteristic scale of around $r_d \approx 147$Mpc that can be used as a standard ruler and to constrain cosmological models. Studies of the BAO feature in the transverse direction provide a measurement of $D_H(z)/r_d = c/H(z)r_d$, with the comoving angular diameter distance \cite{Hogg:2020ktc,Martinelli:2020hud}:
\begin{equation}
D_M= \int_0^z\frac{c \, dz'}{H(z')} \ .
\end{equation}
The angular diameter distance $D_A=D_M/(1+z)$ and the quantity $D_V(z)/r_d$ with
\begin{equation}
D_V(z) \equiv [ z D_H(z) D_M^2(z) ]^{1/3}
\end{equation}
are a combination of the BAO peak coordinates above. The surveys provide the values of the measurements at some effective redshift. We employ the BAO data points collected in \cite{Benisty:2020otr} from \cite{Percival:2009xn,Beutler:2011hx,Busca:2012bu,Anderson:2012sa,Seo:2012xy,Ross:2014qpa,Tojeiro:2014eea,Bautista:2017wwp,deCarvalho:2017xye,Ata:2017dya,Abbott:2017wcz,Molavi:2019mlh}, in the redshift range $0.106 < z < 2.34$. The BAO scale is set by the sound horizon at the drag epoch $z_d \approx 1060$ when photons and baryons decouple. In our analysis we used $r_d$ as independent parameter.

The BAO data represent the absolute distance measurements in the Universe. From the measurements of correlation function or power spectrum of large scale structure, we can use the BAO signal to estimate the distance scales at different redshifts. In practice, the BAO data are analyzed based on a fiducial cosmology and the sound horizon at drag epoch. We calculate the redshift of the drag epoch as:
\begin{equation}
    z_{d}=\frac{1291\omega_{m}^{0.251}}{1+0.659\omega_{m}^{0.828}}[1+b_{1}\omega_{b}^{b_{2}}],
\end{equation}
where
\begin{eqnarray}
    b_{1}&=&0.313\omega_{m}^{-0.419}[1+0.607\omega_{m}^{0.674}], \\
    b_{2}&=&0.238\omega_{m}^{0.223}.
\end{eqnarray}
The uncorrelation of this data set yields the corresponding $\chi^2$:
\begin{equation}
\chi^2_{BAO} = \sum_{i=1}^{17}\left(\frac{D_{i} - D_{pred}(z_i)}{\Delta D_i}\right)^2,
\end{equation}
where $D_{i}$ is the observed distant module rates at redshift $z_{i}$ ($i=1,...,N$) and $D_{pred}$ is the predicted one from the model.

\subsection{CMB distant priors}
We use the CMB distant priors that based on the latest CMB measurements \citep{Aghanim:2018eyx}. Its contribution in the likelihood analysis is expressed in terms of the compressed form with CMB shift parameters:
\begin{eqnarray}
    R &\equiv&\sqrt{\Omega_{m}H_{0}^{2}}r(z_{*})/c, \\
    l_{a} &\equiv& \pi r(z_{*})/r_{s}(z_{*}),
\end{eqnarray}
where $r_{s}(z)$ is the comoving sound horizon at redshift $z$, and $z_{*}$ is the redshift to the photon-decoupling surface. These two CMB shift parameters together with $\omega_{b}=\Omega_{b}h^{2}$ and spectral index of the primordial power spectrum $n_{s}$ can give an efficient summary of CMB data for the dark energy constraints.

The comoving sound horizon is given by
\begin{equation}
    r_{s}(z) =  \frac{c}{H_{0}}\int_{0}^{a}\frac{da'}{\sqrt{3(1+\bar{R_{b}}a')a'^{4}E^{2}(z')}}.
\end{equation}
The radiation term in the expression of $E(z)$ for the CMB data analysis should not be ignored. It can be determined by the matter-radiation equality relation $\Omega_{r}=\Omega_{m}/(1+z_{\text{eq}})$, and $z_{\text{eq}}=2.5\times10^{4}\omega_{m}(T_{\text{CMB}}/2.7\text{K})^{-4}$, where $\omega_{m}=\Omega_{m}h^{2}$. The sound speed is $c_{s}=1/\sqrt{3(1+\bar{R_{b}}a)}$, with $\bar{R_{b}}a=3\rho_{b}/(4\rho_{r})$, and $\bar{R_{b}}=31500w_{b}(T_{\text{CMB}}/2.7\text{K})^{-4}$. We assume the CMB temperature $T_{\text{CMB}}=2.7255\text{K}$. The redshift $z_{*}$ can be calculated by the fitting formula:
\begin{equation}
    z_{*}=1048[1+0.00124\omega_{b}^{-0.738}][1+g_{1}\omega_{m}^{g_{2}}],
\end{equation}
where
\begin{equation}
    g_{1}=\frac{0.0783\omega_{b}^{-0.238}}{1+39.5\omega_{b}^{0.763}}, \qquad g_{2}=\frac{0.560}{1+21.1\omega_{b}^{1.81}}.
\end{equation}

\subsection{Direct Detection of the Hubble parameter}
We include the latest measurement of the Hubble parameter:
\begin{equation}
H_0 = (73.2 \pm 1.3) \text{km/s/Mpc}
\end{equation}
reported by \cite{Riess:2020fzl}. The measurement presents an expanded sample of 75 Milky Way Cepheids with Hubble Space Telescope (HST) photometry and Gaia EDR3 parallaxes which uses the extragalactic distance ladder in order to recalibrate and refine the determination of the Hubble constant.  {The combination is related via the relation}
\begin{equation}
\chi^2_{Hub} = \left(\frac{H_0 - 73.2}{1.3}\right)^2.
\end{equation}
The $\chi^2_{Hub}$ estimates the deviation from the latest measurement of the Hubble constant.

\subsection{Joint analysis and model selection}
In order to perform a joint statistical analysis of $4$ cosmological probes we need to use the total likelihood function, consequently the  $\chi^2_{\text{tot}}$ expression is given by
\begin{equation}
\chi_{\text{tot}}^2 = \chi_{BBN}^2 + \chi_{CMB}^2 + \chi_{H}^2 + \chi_{SC}^2 + \chi_{BAO}^2 + \chi^2_{Hub}.
\end{equation}
In order to perform a joint statistical analysis of these four cosmological probes we need to use the total likelihood function. Regarding the problem of likelihood maximization, we use an affine-invariant Markov Chain Monte Carlo sampler \cite{ForemanMackey:2012ig}, as it is implemented within the open-source packaged $Polychord$ \cite{Handley:2015fda} with the $GetDist$ package \cite{Lewis:2019xzd} to present the results. The prior we choose is with a uniform distribution, where $\Omega_{m} \in [0.;1.]$, $\Omega_{\Lambda}\in[0.;1 - \Omega_{m}]$, $H_0\in [50;100]$Km/sec/Mpc. For the EiBI parameter $\epsilon$  we set  the range $|\kappa^2 \epsilon|\in[0.;10^{-33}]$.

\begin{figure*}[t!]
 	\centering
\includegraphics[width=1\textwidth]{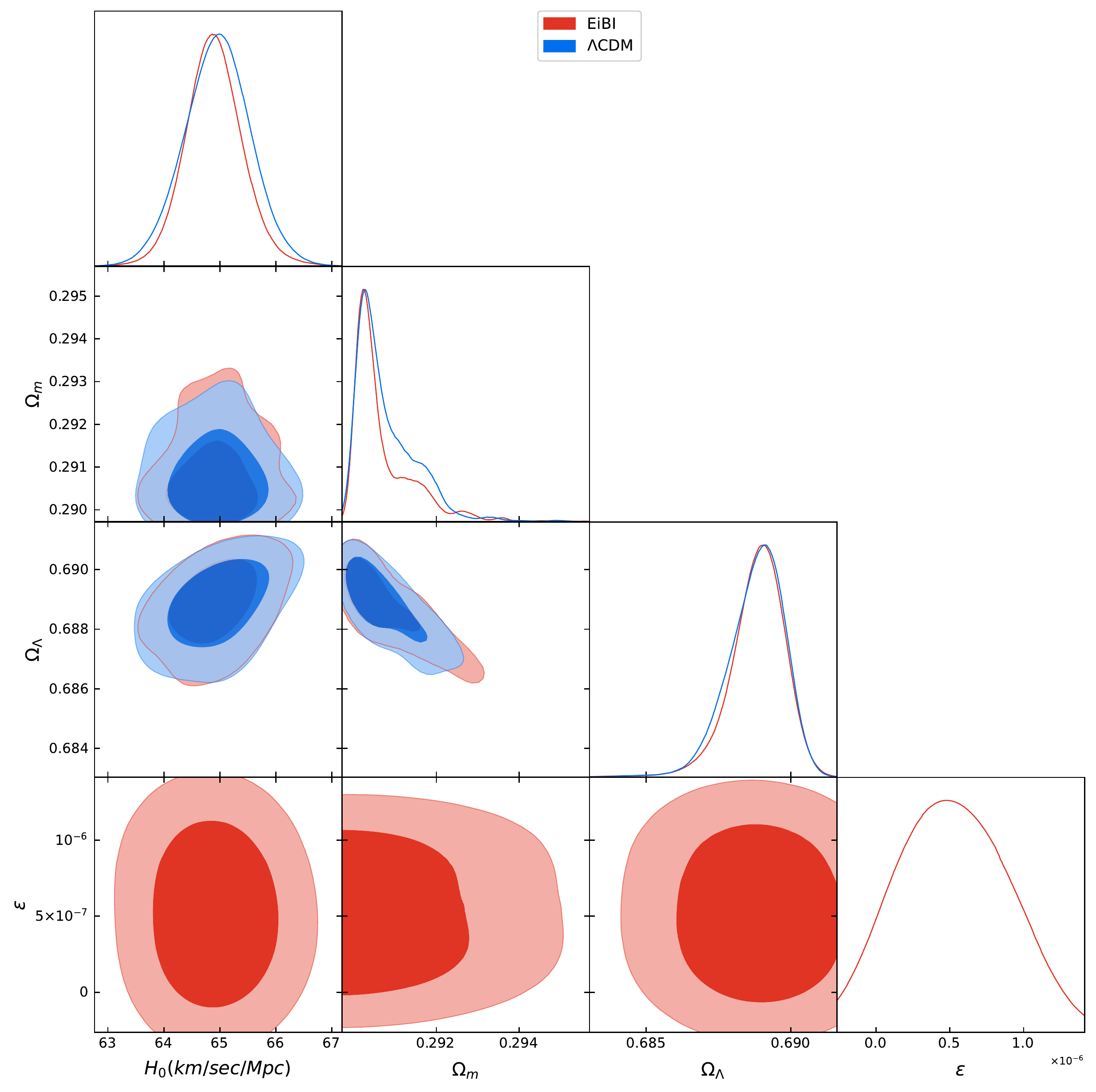}
\begin{tabular}{| c | c | c | c | c | c | c |}
\hline\hline
Model & $H_0 (Km/sec/Mpc)$ & $\Omega_m$ & $\Omega_\Lambda$ & $\epsilon (10^{-8})$ & $ \Delta B_{ij}$\\
\hline\hline
EiBI & $64.99 \pm 0.556$ & $0.291 \pm 0.0074$ & $0.689 \pm 0.0084$ & $\left(5.76 \pm 3.53\right)$  &  $0.51$ \\
\hline
$\Lambda$CDM &  $64.94 \pm 0.525$ & $0.291 \pm 0.008$  & $0.6891 \pm 0.00865$ & - & - \\
\hline\hline
\end{tabular}
\caption{\it{The posterior distribution for the simplest case of the EiBI and for $\Lambda$CDM model (blue curve). The ratios of the matter density $\Omega_m$, dark energy $\Omega_\Lambda$. The final results for cosmological parameters for the EiBI and the $\Lambda$CDM models are summarized in the table. In order to compare the models we calculate the Bayes factor.}}
\label{fig:fit}
\end{figure*}

\begin{figure}[t!]
\centering
\includegraphics[width=1\textwidth]{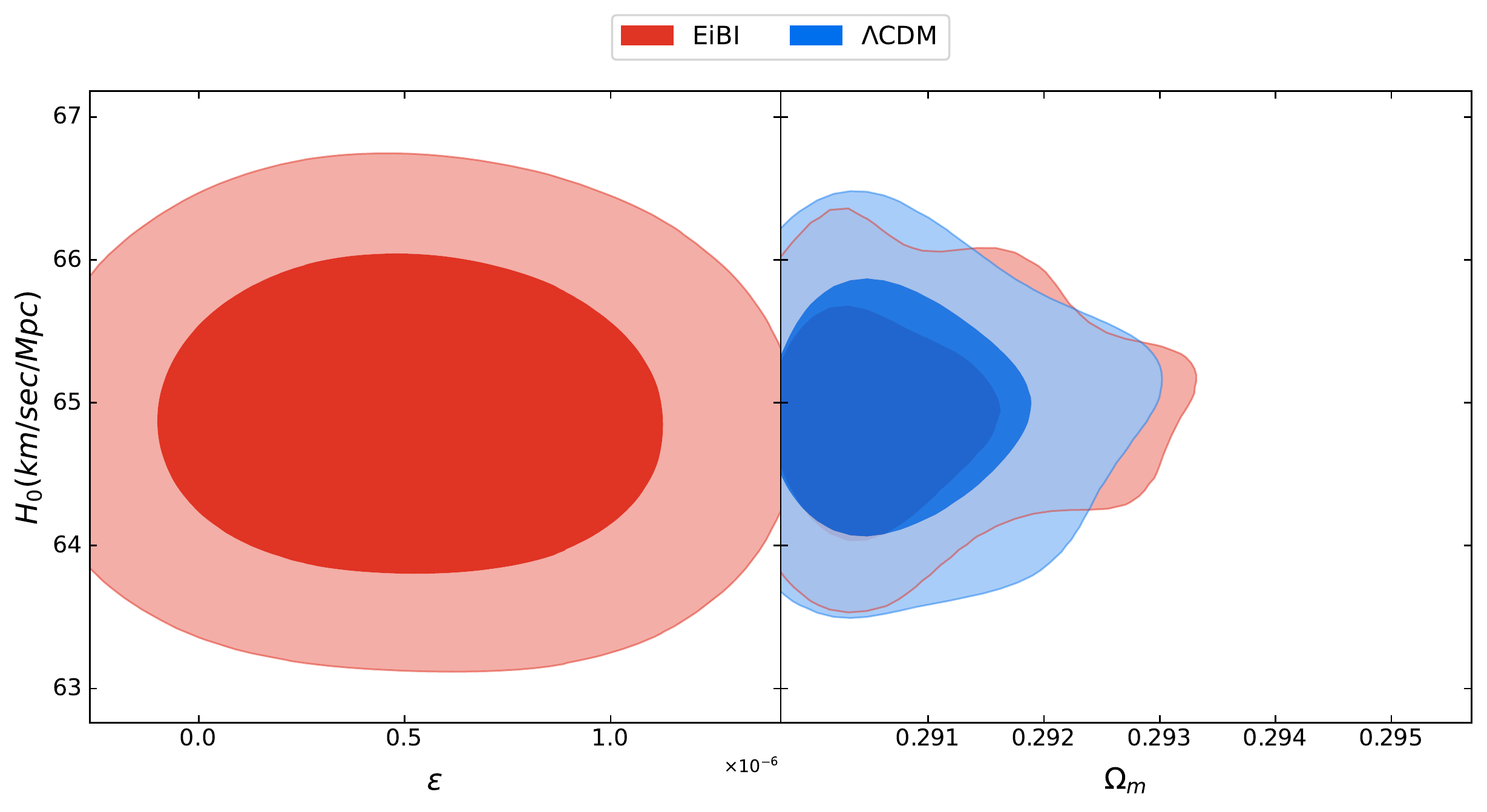}
\caption{The posterior distribution for the Hubble parameter $H_0$ vs. the EiBI parameter $\epsilon$.}
\label{fig:fitSmall}
\end{figure}

The posterior distribution of $H_0$ vs. the EiBI parameter $\epsilon$ is presented in Fig. \ref{fig:fit} and \ref{fig:fitSmall}. The Hubble parameter is $H_0 = 64.99 \pm 0.556$km/sec/Mpc which is in between the Planck estimation of the Hubble function \cite{Aghanim:2018eyx} and the latest SH0ES measurement \cite{Riess:2020fzl}. The dark matter component is $\Omega_m=0.291 \pm 0.0074$ and the dark energy component is being $\Omega_{\Lambda}=0.689 \pm 0.0084$. The fit for the EiBI parameter gives $\kappa^2 \epsilon = \left(4.764 \pm 3.074\right) \cdot 10^{-33}$. We point out that the value of $\epsilon = 0$ (which corresponds to GR) is very close to the mean value. Therefore, for $\vert \epsilon \vert \lesssim \left(5.76 \pm 3.53\right) \cdot 10^{-8}$m$^2$ the theory is able to fit to the data while being able to successfully remove the Big Bang singularity. We point out again that this improves by several orders of magnitude the previous strongest constraints for EiBI gravity's parameter as obtained from particle physics experiments \cite{Jimenez:2021sqd,Latorre:2017uve,Delhom:2019wir}.

We finally point out that from the Bayes factor difference between the models that reads $\Delta B_{ij} = 0.51$ we get an indistinguishable difference for the $\Lambda$CDM model. Therefore, EiBI gravity with the constraint above for its parameter when coupled to a scalar yields very similar properties for the $\Lambda$CDM model, but prevents the initial singularity.

\section{Conclusion and discussion} \label{sec:V}

In this paper we have analyzed homogeneous and isotropic cosmological solutions in the context of Eddington-inspired Born-Infeld gravity coupled to a single scalar field. In the massless case we discussed the existence and properties of such solutions by direct resolution of the modified gravitational field equations, which depend on the sign of the EiBI parameter. For the bouncing solutions ($s=-1$) one finds that at the maximum achievable density of the scalar field the expansion factor gets to a minimum non-vanishing value (vanishing Hubble factor) undergoing a transition from a contracting phase to an expanding one (or vice versa). In the loitering solutions ($s=+1$) the cosmological evolution starts from an asymptotically Minkowski past with a minimum constant value of the expansion factor, such that at a given time a canonical cosmological expansion phase is triggered. These results are consistent with the expectancy raised in the seminal paper \cite{banados} and the numerical evidence found in more recent works \cite{BeltranJimenez:2017uwv} when considering perfect fluids as the matter source.

Though the massless case was known to have only two types of solutions, namely, bouncing and loitering, some kind of instability was expected for the loitering case beyond the massless scenario. Until now it was not known if they would lead to singular solutions of if they would remain nonsingular. Our analysis puts forward that the model is absolutely robust against singularities and that the loitering case gives rise to a new type of  solutions in which an unstable Minkowskian phase (almost constant expansion factor) is possible in between a contraction phase and an expansion phase. The duration of this Minkowskian period (see Fig. \ref{fig:BaL}) depends on model parameters and the initial conditions. No singular solutions are found in our analysis, neither in the massless nor in the massive case. All the nonsingular solutions recover the late-time GR cosmological evolution. 

Getting into the details, for the $s=-1$ (bouncing) branch we found that for low masses the solutions closely resemble the behaviour of those of the massless case. However, for masses above a certain threshold the parametric representation of the Hubble function versus the energy density develops a kind of fish-like asymmetric fins at certain time in the cosmological evolution, a feature not present in the massless case (compare Fig. \ref{fig:PS} with Figs. \ref{fig:BHrhoS} and \ref{fig:BHrhoL}). The $s=+1$ (loitering) branch also exhibits unusual features, showing that during the loitering (almost Minkowskian) phase the energy density can grow significantly  while the Hubble function remains close to zero and positive (see Fig.  \ref{fig:BHrhoSL}). This  represents a kind of intermediate state in a continuous transition from a contracting phase to a final expansion one. This behaviour is triggered by the energy density stored in the massive scalar field, whose drop marks the end of the loitering phase and the beginning of the expansion phase (see Figs. \ref{fig:BrhoLI} and \ref{fig:BrhoLII}).

To conclude, the results obtained in this paper further support the effectiveness of EiBI gravity to remove singularities in cosmological and astrophysical scenarios when coupled to different matter sources. In the present scenario with scalar fields this can be done while being consistent with the latest cosmological observations provided that the EiBI parameter is kept roughly within $\left|\epsilon\right| \lesssim  5 \cdot 10^{-8} \text{ m}^2$. One important aspect of the obtained solutions not tackled here is whether they are stable under tensorial perturbations or not. We point out that in the present case the equation of state is of the form $\omega=\omega(t)$ since the energy density and the pressure of the scalar field (in the massive case) are not trivially related. In turn this should have an impact on the behaviour of the quantities prone to the development of instabilities in the tensorial perturbations equation, such as the sound speed \cite{BeltranJimenez:2017uwv,EscamillaRivera:2012vz}. Therefore, one would need to carry out such an analysis in order to further support the feasibility of this theory and their associated solutions to replace the Big Bang singularity by observationally viable nonsingular cosmologies. \\

\acknowledgments

D.B. gratefully acknowledges the support from the Blavatnik and the Rothschild fellowship. G.J.O. is no longer funded by the Ramon y Cajal contract RYC-2013-13019 (Spain) because he became permanent. D.R.-G. is funded by the \emph{Atracci\'on de Talento Investigador} programme of the Comunidad de Madrid (Spain) No. 2018-T1/TIC-10431, and~acknowledges further support from the Ministerio de Ciencia, Innovaci\'on y Universidades (Spain) project No. PID2019-108485GB-I00/AEI/10.13039/501100011033, and~the FCT projects No. PTDC/FIS-PAR/31938/2017 and PTDC/FIS-OUT/29048/2017.  This work is supported by the Spanish Grant FIS2017-84440-C2- 1-P funded by MCIN/AEI/ 10.13039/501100011033  “ERDF A way of making Europe”, Grant PID2020-116567GB-C21 funded by MCIN/AEI/10.13039/501100011033, the project PROMETEO/2020/079 (Generalitat Valenciana), the~project i-COOPB20462 (CSIC) and the Edital 006/2018 PRONEX (FAPESQ-PB/CNPQ, Brazil, Grant 0015/2019). This article is based on work from COST Action CA18108, supported by COST (European Cooperation in Science and Technology

\bibliography{ref}

\begin{thebibliography}{-------}
\providecommand{\natexlab}[1]{#1}

\bibitem[Amendola \em{et~al.}(2013)Amendola et~al.]{Amendola:2012ys}
Amendola, L.; others.
\newblock {Cosmology and fundamental physics with the Euclid satellite}.
\newblock {\em Living Rev. Rel.} {\bf 2013}, {\em 16},~6,
  \href{http://xxx.lanl.gov/abs/1206.1225}{{\normalfont
  [arXiv:astro-ph.CO/1206.1225]}}.
\newblock
  doi:{\changeurlcolor{black}\href{https://doi.org/10.12942/lrr-2013-6}{\detokenize{10.12942/lrr-2013-6}}}.

\bibitem[Aghanim \em{et~al.}(2020)Aghanim et~al.]{Aghanim:2018eyx}
Aghanim, N.; others.
\newblock {Planck 2018 results. VI. Cosmological parameters}.
\newblock {\em Astron. Astrophys.} {\bf 2020}, {\em 641},~A6,
  \href{http://xxx.lanl.gov/abs/1807.06209}{{\normalfont
  [arXiv:astro-ph.CO/1807.06209]}}.
\newblock
  doi:{\changeurlcolor{black}\href{https://doi.org/10.1051/0004-6361/201833910}{\detokenize{10.1051/0004-6361/201833910}}}.

\bibitem[Starobinsky(1979)]{Starobinsky:1979ty}
Starobinsky, A.A.
\newblock {Spectrum of relict gravitational radiation and the early state of
  the universe}.
\newblock {\em JETP Lett.} {\bf 1979}, {\em 30},~682--685.
\newblock [,767(1979)].

\bibitem[Starobinsky(1980)]{Starobinsky:1980te}
Starobinsky, A.A.
\newblock {A New Type of Isotropic Cosmological Models Without Singularity}.
\newblock {\em Phys. Lett.} {\bf 1980}, {\em 91B},~99--102.
\newblock 771 (1980),
  doi:{\changeurlcolor{black}\href{https://doi.org/10.1016/0370-2693(80)90670-X}{\detokenize{10.1016/0370-2693(80)90670-X}}}.

\bibitem[Guth(1981)]{Guth:1980zm}
Guth, A.H.
\newblock {The Inflationary Universe: A Possible Solution to the Horizon and
  Flatness Problems}.
\newblock {\em Phys. Rev.} {\bf 1981}, {\em D23},~347--356.
\newblock [Adv. Ser. Astrophys. Cosmol.3,139(1987)],
  doi:{\changeurlcolor{black}\href{https://doi.org/10.1103/PhysRevD.23.347}{\detokenize{10.1103/PhysRevD.23.347}}}.

\bibitem[Albrecht and Steinhardt(1982)]{Albrecht:1982wi}
Albrecht, A.; Steinhardt, P.J.
\newblock {Cosmology for Grand Unified Theories with Radiatively Induced
  Symmetry Breaking}.
\newblock {\em Phys. Rev. Lett.} {\bf 1982}, {\em 48},~1220--1223.
\newblock [Adv. Ser. Astrophys. Cosmol.3,158(1987)],
  doi:{\changeurlcolor{black}\href{https://doi.org/10.1103/PhysRevLett.48.1220}{\detokenize{10.1103/PhysRevLett.48.1220}}}.

\bibitem[Mukhanov and Chibisov(1981)]{Mukhanov:1981xt}
Mukhanov, V.F.; Chibisov, G.V.
\newblock {Quantum Fluctuations and a Nonsingular Universe}.
\newblock {\em JETP Lett.} {\bf 1981}, {\em 33},~532--535.
\newblock [Pisma Zh. Eksp. Teor. Fiz.33,549(1981)].

\bibitem[Guth and Pi(1982)]{Guth:1982ec}
Guth, A.H.; Pi, S.Y.
\newblock {Fluctuations in the New Inflationary Universe}.
\newblock {\em Phys. Rev. Lett.} {\bf 1982}, {\em 49},~1110--1113.
\newblock
  doi:{\changeurlcolor{black}\href{https://doi.org/10.1103/PhysRevLett.49.1110}{\detokenize{10.1103/PhysRevLett.49.1110}}}.

\bibitem[Linde(1982)]{Linde:1981mu}
Linde, A.D.
\newblock {A New Inflationary Universe Scenario: A Possible Solution of the
  Horizon, Flatness, Homogeneity, Isotropy and Primordial Monopole Problems}.
\newblock {\em Phys. Lett.} {\bf 1982}, {\em 108B},~389--393.
\newblock [Adv. Ser. Astrophys. Cosmol.3,149(1987)],
  doi:{\changeurlcolor{black}\href{https://doi.org/10.1016/0370-2693(82)91219-9}{\detokenize{10.1016/0370-2693(82)91219-9}}}.

\bibitem[Barrow and Cotsakis(1988)]{Barrow:1988xh}
Barrow, J.D.; Cotsakis, S.
\newblock {Inflation and the Conformal Structure of Higher Order Gravity
  Theories}.
\newblock {\em Phys. Lett.} {\bf 1988}, {\em B214},~515--518.
\newblock
  doi:{\changeurlcolor{black}\href{https://doi.org/10.1016/0370-2693(88)90110-4}{\detokenize{10.1016/0370-2693(88)90110-4}}}.

\bibitem[Barrow(1988)]{Barrow:1988xi}
Barrow, J.D.
\newblock {The Premature Recollapse Problem in Closed Inflationary Universes}.
\newblock {\em Nucl. Phys. B} {\bf 1988}, {\em 296},~697--709.
\newblock
  doi:{\changeurlcolor{black}\href{https://doi.org/10.1016/0550-3213(88)90040-5}{\detokenize{10.1016/0550-3213(88)90040-5}}}.

\bibitem[Elizalde \em{et~al.}(2008)Elizalde, Nojiri, Odintsov, Saez-Gomez, and
  Faraoni]{Elizalde:2008yf}
Elizalde, E.; Nojiri, S.; Odintsov, S.D.; Saez-Gomez, D.; Faraoni, V.
\newblock {Reconstructing the universe history, from inflation to acceleration,
  with phantom and canonical scalar fields}.
\newblock {\em Phys. Rev. D} {\bf 2008}, {\em 77},~106005,
  \href{http://xxx.lanl.gov/abs/0803.1311}{{\normalfont
  [arXiv:hep-th/0803.1311]}}.
\newblock
  doi:{\changeurlcolor{black}\href{https://doi.org/10.1103/PhysRevD.77.106005}{\detokenize{10.1103/PhysRevD.77.106005}}}.

\bibitem[Ratra and Peebles(1988)]{Ratra:1987rm}
Ratra, B.; Peebles, P.J.E.
\newblock {Cosmological Consequences of a Rolling Homogeneous Scalar Field}.
\newblock {\em Phys. Rev.} {\bf 1988}, {\em D37},~3406.
\newblock
  doi:{\changeurlcolor{black}\href{https://doi.org/10.1103/PhysRevD.37.3406}{\detokenize{10.1103/PhysRevD.37.3406}}}.

\bibitem[Caldwell \em{et~al.}(1998)Caldwell, Dave, and
  Steinhardt]{Caldwell:1997ii}
Caldwell, R.R.; Dave, R.; Steinhardt, P.J.
\newblock {Cosmological imprint of an energy component with general equation of
  state}.
\newblock {\em Phys. Rev. Lett.} {\bf 1998}, {\em 80},~1582--1585,
  \href{http://xxx.lanl.gov/abs/astro-ph/9708069}{{\normalfont
  [arXiv:astro-ph/astro-ph/9708069]}}.
\newblock
  doi:{\changeurlcolor{black}\href{https://doi.org/10.1103/PhysRevLett.80.1582}{\detokenize{10.1103/PhysRevLett.80.1582}}}.

\bibitem[Kehayias and Scherrer(2019)]{Kehayias:2019gir}
Kehayias, J.; Scherrer, R.J.
\newblock {New generic evolution for $k$ -essence dark energy with $w \approx
  -1$}.
\newblock {\em Phys. Rev.} {\bf 2019}, {\em D100},~023525,
  \href{http://xxx.lanl.gov/abs/1905.05628}{{\normalfont
  [arXiv:gr-qc/1905.05628]}}.
\newblock
  doi:{\changeurlcolor{black}\href{https://doi.org/10.1103/PhysRevD.100.023525}{\detokenize{10.1103/PhysRevD.100.023525}}}.

\bibitem[Oikonomou and Chatzarakis(2020)]{Oikonomou:2019muq}
Oikonomou, V.K.; Chatzarakis, N.
\newblock {The Phase Space of $k$-Essence $f(R)$ Gravity Theory}.
\newblock {\em Nucl. Phys.} {\bf 2020}, {\em B956},~115023,
  \href{http://xxx.lanl.gov/abs/1905.01904}{{\normalfont
  [arXiv:gr-qc/1905.01904]}}.
\newblock
  doi:{\changeurlcolor{black}\href{https://doi.org/10.1016/j.nuclphysb.2020.115023}{\detokenize{10.1016/j.nuclphysb.2020.115023}}}.

\bibitem[Chakraborty \em{et~al.}(2019)Chakraborty, Ghosh, and
  Banerjee]{Chakraborty:2019swx}
Chakraborty, A.; Ghosh, A.; Banerjee, N.
\newblock {Dynamical systems analysis of a k -essence model}.
\newblock {\em Phys. Rev.} {\bf 2019}, {\em D99},~103513,
  \href{http://xxx.lanl.gov/abs/1904.10149}{{\normalfont
  [arXiv:gr-qc/1904.10149]}}.
\newblock
  doi:{\changeurlcolor{black}\href{https://doi.org/10.1103/PhysRevD.99.103513}{\detokenize{10.1103/PhysRevD.99.103513}}}.

\bibitem[Babichev \em{et~al.}(2018)Babichev, Ramazanov, and
  Vikman]{Babichev:2018twg}
Babichev, E.; Ramazanov, S.; Vikman, A.
\newblock {Recovering $P(X)$ from a canonical complex field} {\bf 2018}.
\newblock  \href{http://xxx.lanl.gov/abs/1807.10281}{{\normalfont
  [arXiv:gr-qc/1807.10281]}}.
\newblock [JCAP1811,023(2018)],
  doi:{\changeurlcolor{black}\href{https://doi.org/10.1088/1475-7516/2018/11/023}{\detokenize{10.1088/1475-7516/2018/11/023}}}.

\bibitem[Zlatev \em{et~al.}(1999)Zlatev, Wang, and Steinhardt]{Zlatev:1998tr}
Zlatev, I.; Wang, L.M.; Steinhardt, P.J.
\newblock {Quintessence, cosmic coincidence, and the cosmological constant}.
\newblock {\em Phys. Rev. Lett.} {\bf 1999}, {\em 82},~896--899,
  \href{http://xxx.lanl.gov/abs/astro-ph/9807002}{{\normalfont
  [arXiv:astro-ph/astro-ph/9807002]}}.
\newblock
  doi:{\changeurlcolor{black}\href{https://doi.org/10.1103/PhysRevLett.82.896}{\detokenize{10.1103/PhysRevLett.82.896}}}.

\bibitem[Caldwell(2002)]{Caldwell:1999ew}
Caldwell, R.R.
\newblock {A Phantom menace?}
\newblock {\em Phys. Lett.} {\bf 2002}, {\em B545},~23--29,
  \href{http://xxx.lanl.gov/abs/astro-ph/9908168}{{\normalfont
  [arXiv:astro-ph/astro-ph/9908168]}}.
\newblock
  doi:{\changeurlcolor{black}\href{https://doi.org/10.1016/S0370-2693(02)02589-3}{\detokenize{10.1016/S0370-2693(02)02589-3}}}.

\bibitem[Chiba \em{et~al.}(2000)Chiba, Okabe, and Yamaguchi]{Chiba:1999ka}
Chiba, T.; Okabe, T.; Yamaguchi, M.
\newblock {Kinetically driven quintessence}.
\newblock {\em Phys. Rev.} {\bf 2000}, {\em D62},~023511,
  \href{http://xxx.lanl.gov/abs/astro-ph/9912463}{{\normalfont
  [arXiv:astro-ph/astro-ph/9912463]}}.
\newblock
  doi:{\changeurlcolor{black}\href{https://doi.org/10.1103/PhysRevD.62.023511}{\detokenize{10.1103/PhysRevD.62.023511}}}.

\bibitem[Bento \em{et~al.}(2002)Bento, Bertolami, and Sen]{Bento:2002ps}
Bento, M.C.; Bertolami, O.; Sen, A.A.
\newblock {Generalized Chaplygin gas, accelerated expansion and dark energy
  matter unification}.
\newblock {\em Phys. Rev.} {\bf 2002}, {\em D66},~043507,
  \href{http://xxx.lanl.gov/abs/gr-qc/0202064}{{\normalfont
  [arXiv:gr-qc/gr-qc/0202064]}}.
\newblock
  doi:{\changeurlcolor{black}\href{https://doi.org/10.1103/PhysRevD.66.043507}{\detokenize{10.1103/PhysRevD.66.043507}}}.

\bibitem[Tsujikawa(2013)]{Tsujikawa:2013fta}
Tsujikawa, S.
\newblock {Quintessence: A Review}.
\newblock {\em Class. Quant. Grav.} {\bf 2013}, {\em 30},~214003,
  \href{http://xxx.lanl.gov/abs/1304.1961}{{\normalfont
  [arXiv:gr-qc/1304.1961]}}.
\newblock
  doi:{\changeurlcolor{black}\href{https://doi.org/10.1088/0264-9381/30/21/214003}{\detokenize{10.1088/0264-9381/30/21/214003}}}.

\bibitem[Hu \em{et~al.}(2000)Hu, Barkana, and Gruzinov]{Hu:2000ke}
Hu, W.; Barkana, R.; Gruzinov, A.
\newblock {Cold and fuzzy dark matter}.
\newblock {\em Phys. Rev. Lett.} {\bf 2000}, {\em 85},~1158--1161,
  \href{http://xxx.lanl.gov/abs/astro-ph/0003365}{{\normalfont
  [astro-ph/0003365]}}.
\newblock
  doi:{\changeurlcolor{black}\href{https://doi.org/10.1103/PhysRevLett.85.1158}{\detokenize{10.1103/PhysRevLett.85.1158}}}.

\bibitem[Anagnostopoulos \em{et~al.}(2019)Anagnostopoulos, Benisty, Basilakos,
  and Guendelman]{Anagnostopoulos:2019myt}
Anagnostopoulos, F.K.; Benisty, D.; Basilakos, S.; Guendelman, E.I.
\newblock {Dark energy and dark matter unification from dynamical space time:
  observational constraints and cosmological implications}.
\newblock {\em JCAP} {\bf 2019}, {\em 1906},~003,
  \href{http://xxx.lanl.gov/abs/1904.05762}{{\normalfont
  [arXiv:gr-qc/1904.05762]}}.
\newblock
  doi:{\changeurlcolor{black}\href{https://doi.org/10.1088/1475-7516/2019/06/003}{\detokenize{10.1088/1475-7516/2019/06/003}}}.

\bibitem[Benisty \em{et~al.}(2019)Benisty, Guendelman, and
  Haba]{Benisty:2018oyy}
Benisty, D.; Guendelman, E.; Haba, Z.
\newblock {Unification of dark energy and dark matter from diffusive
  cosmology}.
\newblock {\em Phys. Rev.} {\bf 2019}, {\em D99},~123521,
  \href{http://xxx.lanl.gov/abs/1812.06151}{{\normalfont
  [arXiv:gr-qc/1812.06151]}}.
\newblock
  doi:{\changeurlcolor{black}\href{https://doi.org/10.1103/PhysRevD.99.123521}{\detokenize{10.1103/PhysRevD.99.123521}}}.

\bibitem[Benisty and Guendelman(2018)]{Benisty:2018qed}
Benisty, D.; Guendelman, E.I.
\newblock {Unified dark energy and dark matter from dynamical spacetime}.
\newblock {\em Phys. Rev.} {\bf 2018}, {\em D98},~023506,
  \href{http://xxx.lanl.gov/abs/1802.07981}{{\normalfont
  [arXiv:gr-qc/1802.07981]}}.
\newblock
  doi:{\changeurlcolor{black}\href{https://doi.org/10.1103/PhysRevD.98.023506}{\detokenize{10.1103/PhysRevD.98.023506}}}.

\bibitem[Benisty and Guendelman(2017)]{Benisty:2017eqh}
Benisty, D.; Guendelman, E.I.
\newblock {Interacting Diffusive Unified Dark Energy and Dark Matter from
  Scalar Fields}.
\newblock {\em Eur. Phys. J.} {\bf 2017}, {\em C77},~396,
  \href{http://xxx.lanl.gov/abs/1701.08667}{{\normalfont
  [arXiv:gr-qc/1701.08667]}}.
\newblock
  doi:{\changeurlcolor{black}\href{https://doi.org/10.1140/epjc/s10052-017-4939-x}{\detokenize{10.1140/epjc/s10052-017-4939-x}}}.

\bibitem[Senovilla and Garfinkle(2015)]{Senovilla:2014gza}
Senovilla, J.M.M.; Garfinkle, D.
\newblock {The 1965 Penrose singularity theorem}.
\newblock {\em Class. Quant. Grav.} {\bf 2015}, {\em 32},~124008,
  \href{http://xxx.lanl.gov/abs/1410.5226}{{\normalfont
  [arXiv:gr-qc/1410.5226]}}.
\newblock
  doi:{\changeurlcolor{black}\href{https://doi.org/10.1088/0264-9381/32/12/124008}{\detokenize{10.1088/0264-9381/32/12/124008}}}.

\bibitem[De~Felice and Tsujikawa(2010)]{DeFelice:2010aj}
De~Felice, A.; Tsujikawa, S.
\newblock {f(R) theories}.
\newblock {\em Living Rev. Rel.} {\bf 2010}, {\em 13},~3,
  \href{http://xxx.lanl.gov/abs/1002.4928}{{\normalfont
  [arXiv:gr-qc/1002.4928]}}.
\newblock
  doi:{\changeurlcolor{black}\href{https://doi.org/10.12942/lrr-2010-3}{\detokenize{10.12942/lrr-2010-3}}}.

\bibitem[Capozziello and De~Laurentis(2011)]{CLreview}
Capozziello, S.; De~Laurentis, M.
\newblock {Extended Theories of Gravity}.
\newblock {\em Phys. Rept.} {\bf 2011}, {\em 509},~167--321,
  \href{http://xxx.lanl.gov/abs/1108.6266}{{\normalfont
  [arXiv:gr-qc/1108.6266]}}.
\newblock
  doi:{\changeurlcolor{black}\href{https://doi.org/10.1016/j.physrep.2011.09.003}{\detokenize{10.1016/j.physrep.2011.09.003}}}.

\bibitem[Nojiri \em{et~al.}(2017)Nojiri, Odintsov, and
  Oikonomou]{Nojiri:2017ncd}
Nojiri, S.; Odintsov, S.D.; Oikonomou, V.K.
\newblock {Modified Gravity Theories on a Nutshell: Inflation, Bounce and
  Late-time Evolution}.
\newblock {\em Phys. Rept.} {\bf 2017}, {\em 692},~1--104,
  \href{http://xxx.lanl.gov/abs/1705.11098}{{\normalfont
  [arXiv:gr-qc/1705.11098]}}.
\newblock
  doi:{\changeurlcolor{black}\href{https://doi.org/10.1016/j.physrep.2017.06.001}{\detokenize{10.1016/j.physrep.2017.06.001}}}.

\bibitem[Heisenberg(2019)]{Heisenberg:2018vsk}
Heisenberg, L.
\newblock {A systematic approach to generalisations of General Relativity and
  their cosmological implications}.
\newblock {\em Phys. Rept.} {\bf 2019}, {\em 796},~1--113,
  \href{http://xxx.lanl.gov/abs/1807.01725}{{\normalfont
  [arXiv:gr-qc/1807.01725]}}.
\newblock
  doi:{\changeurlcolor{black}\href{https://doi.org/10.1016/j.physrep.2018.11.006}{\detokenize{10.1016/j.physrep.2018.11.006}}}.

\bibitem[Bull \em{et~al.}(2016)Bull et~al.]{Bull:2015stt}
Bull, P.; others.
\newblock {Beyond $\Lambda$CDM: Problems, solutions, and the road ahead}.
\newblock {\em Phys. Dark Univ.} {\bf 2016}, {\em 12},~56--99,
  \href{http://xxx.lanl.gov/abs/1512.05356}{{\normalfont
  [arXiv:astro-ph.CO/1512.05356]}}.
\newblock
  doi:{\changeurlcolor{black}\href{https://doi.org/10.1016/j.dark.2016.02.001}{\detokenize{10.1016/j.dark.2016.02.001}}}.

\bibitem[Banados and Ferreira(2010)]{banados}
Banados, M.; Ferreira, P.G.
\newblock {Eddington's theory of gravity and its progeny}.
\newblock {\em Phys. Rev. Lett.} {\bf 2010}, {\em 105},~011101,
  \href{http://xxx.lanl.gov/abs/1006.1769}{{\normalfont
  [arXiv:astro-ph.CO/1006.1769]}}.
\newblock [Erratum: Phys.Rev.Lett. 113, 119901 (2014)],
  doi:{\changeurlcolor{black}\href{https://doi.org/10.1103/PhysRevLett.105.011101}{\detokenize{10.1103/PhysRevLett.105.011101}}}.

\bibitem[Alishahiha \em{et~al.}(2004)Alishahiha, Silverstein, and
  Tong]{Alishahiha:2004eh}
Alishahiha, M.; Silverstein, E.; Tong, D.
\newblock {DBI in the sky}.
\newblock {\em Phys. Rev. D} {\bf 2004}, {\em 70},~123505,
  \href{http://xxx.lanl.gov/abs/hep-th/0404084}{{\normalfont
  [hep-th/0404084]}}.
\newblock
  doi:{\changeurlcolor{black}\href{https://doi.org/10.1103/PhysRevD.70.123505}{\detokenize{10.1103/PhysRevD.70.123505}}}.

\bibitem[Liu \em{et~al.}(2012)Liu, Yang, Guo, and Zhong]{Liu:2012rc}
Liu, Y.X.; Yang, K.; Guo, H.; Zhong, Y.
\newblock {Domain Wall Brane in Eddington Inspired Born-Infeld Gravity}.
\newblock {\em Phys. Rev. D} {\bf 2012}, {\em 85},~124053,
  \href{http://xxx.lanl.gov/abs/1203.2349}{{\normalfont
  [arXiv:hep-th/1203.2349]}}.
\newblock
  doi:{\changeurlcolor{black}\href{https://doi.org/10.1103/PhysRevD.85.124053}{\detokenize{10.1103/PhysRevD.85.124053}}}.

\bibitem[Choudhury and Pal(2013)]{Choudhury:2012yh}
Choudhury, S.; Pal, S.
\newblock {DBI Galileon inflation in background SUGRA}.
\newblock {\em Nucl. Phys. B} {\bf 2013}, {\em 874},~85--114,
  \href{http://xxx.lanl.gov/abs/1208.4433}{{\normalfont
  [arXiv:hep-th/1208.4433]}}.
\newblock
  doi:{\changeurlcolor{black}\href{https://doi.org/10.1016/j.nuclphysb.2013.05.010}{\detokenize{10.1016/j.nuclphysb.2013.05.010}}}.

\bibitem[Choudhury and Pal(2015)]{Choudhury:2015yna}
Choudhury, S.; Pal, S.
\newblock {Primordial non-Gaussian features from DBI Galileon inflation}.
\newblock {\em Eur. Phys. J. C} {\bf 2015}, {\em 75},~241,
  \href{http://xxx.lanl.gov/abs/1210.4478}{{\normalfont
  [arXiv:hep-th/1210.4478]}}.
\newblock
  doi:{\changeurlcolor{black}\href{https://doi.org/10.1140/epjc/s10052-015-3452-3}{\detokenize{10.1140/epjc/s10052-015-3452-3}}}.

\bibitem[Harko \em{et~al.}(2013)Harko, Lobo, Mak, and Sushkov]{Harko:2013wka}
Harko, T.; Lobo, F.S.N.; Mak, M.K.; Sushkov, S.V.
\newblock {Structure of neutron, quark and exotic stars in Eddington-inspired
  Born-Infeld gravity}.
\newblock {\em Phys. Rev. D} {\bf 2013}, {\em 88},~044032,
  \href{http://xxx.lanl.gov/abs/1305.6770}{{\normalfont
  [arXiv:gr-qc/1305.6770]}}.
\newblock
  doi:{\changeurlcolor{black}\href{https://doi.org/10.1103/PhysRevD.88.044032}{\detokenize{10.1103/PhysRevD.88.044032}}}.

\bibitem[Wei \em{et~al.}(2015)Wei, Yang, and Liu]{Wei:2014dka}
Wei, S.W.; Yang, K.; Liu, Y.X.
\newblock {Black hole solution and strong gravitational lensing in
  Eddington-inspired Born\textendash{}Infeld gravity}.
\newblock {\em Eur. Phys. J. C} {\bf 2015}, {\em 75},~253,
  \href{http://xxx.lanl.gov/abs/1405.2178}{{\normalfont
  [arXiv:gr-qc/1405.2178]}}.
\newblock [Erratum: Eur.Phys.J.C 75, 331 (2015)],
  doi:{\changeurlcolor{black}\href{https://doi.org/10.1140/epjc/s10052-015-3556-9}{\detokenize{10.1140/epjc/s10052-015-3556-9}}}.

\bibitem[Shaikh(2015)]{Shaikh:2015oha}
Shaikh, R.
\newblock {Lorentzian wormholes in Eddington-inspired Born-Infeld gravity}.
\newblock {\em Phys. Rev. D} {\bf 2015}, {\em 92},~024015,
  \href{http://xxx.lanl.gov/abs/1505.01314}{{\normalfont
  [arXiv:gr-qc/1505.01314]}}.
\newblock
  doi:{\changeurlcolor{black}\href{https://doi.org/10.1103/PhysRevD.92.024015}{\detokenize{10.1103/PhysRevD.92.024015}}}.

\bibitem[Avelino(2016)]{Avelino:2015fve}
Avelino, P.P.
\newblock {Inner Structure of Black Holes in Eddington-inspired Born-Infeld
  gravity: the role of mass inflation}.
\newblock {\em Phys. Rev. D} {\bf 2016}, {\em 93},~044067,
  \href{http://xxx.lanl.gov/abs/1511.03223}{{\normalfont
  [arXiv:gr-qc/1511.03223]}}.
\newblock
  doi:{\changeurlcolor{black}\href{https://doi.org/10.1103/PhysRevD.93.044067}{\detokenize{10.1103/PhysRevD.93.044067}}}.

\bibitem[Prasetyo \em{et~al.}(2018)Prasetyo, Husin, Qauli, Ramadhan, and
  Sulaksono]{Prasetyo:2017hrb}
Prasetyo, I.; Husin, I.; Qauli, A.I.; Ramadhan, H.S.; Sulaksono, A.
\newblock {Neutron stars in the braneworld within the Eddington-inspired
  Born-Infeld gravity}.
\newblock {\em JCAP} {\bf 2018}, {\em 01},~027,
  \href{http://xxx.lanl.gov/abs/1708.04837}{{\normalfont
  [arXiv:astro-ph.CO/1708.04837]}}.
\newblock
  doi:{\changeurlcolor{black}\href{https://doi.org/10.1088/1475-7516/2018/01/027}{\detokenize{10.1088/1475-7516/2018/01/027}}}.

\bibitem[Chen \em{et~al.}(2018)Chen, Bouhmadi-L\'opez, and Chen]{Chen:2017ify}
Chen, C.Y.; Bouhmadi-L\'opez, M.; Chen, P.
\newblock {Black hole solutions in mimetic Born-Infeld gravity}.
\newblock {\em Eur. Phys. J. C} {\bf 2018}, {\em 78},~59,
  \href{http://xxx.lanl.gov/abs/1710.10638}{{\normalfont
  [arXiv:gr-qc/1710.10638]}}.
\newblock
  doi:{\changeurlcolor{black}\href{https://doi.org/10.1140/epjc/s10052-018-5556-z}{\detokenize{10.1140/epjc/s10052-018-5556-z}}}.

\bibitem[Shaikh(2018)]{Shaikh:2018yku}
Shaikh, R.
\newblock {Wormholes with nonexotic matter in Born-Infeld gravity}.
\newblock {\em Phys. Rev. D} {\bf 2018}, {\em 98},~064033,
  \href{http://xxx.lanl.gov/abs/1807.07941}{{\normalfont
  [arXiv:gr-qc/1807.07941]}}.
\newblock
  doi:{\changeurlcolor{black}\href{https://doi.org/10.1103/PhysRevD.98.064033}{\detokenize{10.1103/PhysRevD.98.064033}}}.

\bibitem[Jana \em{et~al.}(2018)Jana, Shaikh, and Sarkar]{Jana:2018knq}
Jana, S.; Shaikh, R.; Sarkar, S.
\newblock {Overcharging black holes and cosmic censorship in Born-Infeld
  gravity}.
\newblock {\em Phys. Rev. D} {\bf 2018}, {\em 98},~124039,
  \href{http://xxx.lanl.gov/abs/1808.09656}{{\normalfont
  [arXiv:gr-qc/1808.09656]}}.
\newblock
  doi:{\changeurlcolor{black}\href{https://doi.org/10.1103/PhysRevD.98.124039}{\detokenize{10.1103/PhysRevD.98.124039}}}.

\bibitem[B\"ohmer and Fiorini(2019)]{Boehmer:2019uxv}
B\"ohmer, C.G.; Fiorini, F.
\newblock {The regular black hole in four dimensional Born\textendash{}Infeld
  gravity}.
\newblock {\em Class. Quant. Grav.} {\bf 2019}, {\em 36},~12LT01,
  \href{http://xxx.lanl.gov/abs/1901.02965}{{\normalfont
  [arXiv:gr-qc/1901.02965]}}.
\newblock
  doi:{\changeurlcolor{black}\href{https://doi.org/10.1088/1361-6382/ab1e8d}{\detokenize{10.1088/1361-6382/ab1e8d}}}.

\bibitem[Delhom \em{et~al.}(2019)Delhom, Macedo, Olmo, and
  Crispino]{Delhom:2019btt}
Delhom, A.; Macedo, C.F.B.; Olmo, G.J.; Crispino, L.C.B.
\newblock {Absorption by black hole remnants in metric-affine gravity}.
\newblock {\em Phys. Rev.} {\bf 2019}, {\em D100},~024016,
  \href{http://xxx.lanl.gov/abs/1906.06411}{{\normalfont
  [arXiv:gr-qc/1906.06411]}}.
\newblock
  doi:{\changeurlcolor{black}\href{https://doi.org/10.1103/PhysRevD.100.024016}{\detokenize{10.1103/PhysRevD.100.024016}}}.

\bibitem[Avelino and Ferreira(2012)]{Avelino:2012ue}
Avelino, P.P.; Ferreira, R.Z.
\newblock {Bouncing Eddington-inspired Born-Infeld cosmologies: an alternative
  to Inflation ?}
\newblock {\em Phys. Rev. D} {\bf 2012}, {\em 86},~041501,
  \href{http://xxx.lanl.gov/abs/1205.6676}{{\normalfont
  [arXiv:astro-ph.CO/1205.6676]}}.
\newblock
  doi:{\changeurlcolor{black}\href{https://doi.org/10.1103/PhysRevD.86.041501}{\detokenize{10.1103/PhysRevD.86.041501}}}.

\bibitem[Beltran~Jimenez \em{et~al.}(2018)Beltran~Jimenez, Heisenberg, Olmo,
  and Rubiera-Garcia]{BeltranJimenez:2017doy}
Beltran~Jimenez, J.; Heisenberg, L.; Olmo, G.J.; Rubiera-Garcia, D.
\newblock {Born\textendash{}Infeld inspired modifications of gravity}.
\newblock {\em Phys. Rept.} {\bf 2018}, {\em 727},~1--129,
  \href{http://xxx.lanl.gov/abs/1704.03351}{{\normalfont
  [arXiv:gr-qc/1704.03351]}}.
\newblock
  doi:{\changeurlcolor{black}\href{https://doi.org/10.1016/j.physrep.2017.11.001}{\detokenize{10.1016/j.physrep.2017.11.001}}}.

\bibitem[Beltr\'an~Jim\'enez and Delhom(2019)]{BeltranJimenez:2019acz}
Beltr\'an~Jim\'enez, J.; Delhom, A.
\newblock {Ghosts in metric-affine higher order curvature gravity}.
\newblock {\em Eur. Phys. J. C} {\bf 2019}, {\em 79},~656,
  \href{http://xxx.lanl.gov/abs/1901.08988}{{\normalfont
  [arXiv:gr-qc/1901.08988]}}.
\newblock
  doi:{\changeurlcolor{black}\href{https://doi.org/10.1140/epjc/s10052-019-7149-x}{\detokenize{10.1140/epjc/s10052-019-7149-x}}}.

\bibitem[Beltrán~Jiménez and Delhom(2020)]{Jimenez:2020dpn}
Beltrán~Jiménez, J.; Delhom, A.
\newblock {Instabilities in metric-affine theories of gravity with higher order
  curvature terms}.
\newblock {\em Eur. Phys. J.} {\bf 2020}, {\em C80},~585,
  \href{http://xxx.lanl.gov/abs/2004.11357}{{\normalfont
  [arXiv:gr-qc/2004.11357]}}.
\newblock
  doi:{\changeurlcolor{black}\href{https://doi.org/10.1140/epjc/s10052-020-8143-z}{\detokenize{10.1140/epjc/s10052-020-8143-z}}}.

\bibitem[Delhom(2020)]{Delhom:2020hkb}
Delhom, A.
\newblock {Minimal coupling in presence of non-metricity and torsion}.
\newblock {\em Eur. Phys. J.} {\bf 2020}, {\em C80},~728,
  \href{http://xxx.lanl.gov/abs/2002.02404}{{\normalfont
  [arXiv:gr-qc/2002.02404]}}.
\newblock
  doi:{\changeurlcolor{black}\href{https://doi.org/10.1140/epjc/s10052-020-8330-y}{\detokenize{10.1140/epjc/s10052-020-8330-y}}}.

\bibitem[Afonso \em{et~al.}(2018)Afonso, Olmo, and
  Rubiera-Garcia]{Afonso:2018bpv}
Afonso, V.I.; Olmo, G.J.; Rubiera-Garcia, D.
\newblock {Mapping Ricci-based theories of gravity into general relativity}.
\newblock {\em Phys. Rev. D} {\bf 2018}, {\em 97},~021503,
  \href{http://xxx.lanl.gov/abs/1801.10406}{{\normalfont
  [arXiv:gr-qc/1801.10406]}}.
\newblock
  doi:{\changeurlcolor{black}\href{https://doi.org/10.1103/PhysRevD.97.021503}{\detokenize{10.1103/PhysRevD.97.021503}}}.

\bibitem[Beltrán~Jiménez \em{et~al.}(2020)Beltrán~Jiménez, De~Andrés, and
  Delhom]{Jimenez:2020iok}
Beltrán~Jiménez, J.; De~Andrés, D.; Delhom, A.
\newblock {Anisotropic deformations in a class of projectively-invariant
  metric-affine theories of gravity}.
\newblock {\em Class. Quant. Grav.} {\bf 2020}, {\em 37},~225013,
  \href{http://xxx.lanl.gov/abs/2006.07406}{{\normalfont
  [arXiv:gr-qc/2006.07406]}}.
\newblock
  doi:{\changeurlcolor{black}\href{https://doi.org/10.1088/1361-6382/abb923}{\detokenize{10.1088/1361-6382/abb923}}}.

\bibitem[Jana \em{et~al.}(2018)Jana, Chakravarty, and Mohanty]{Jana:2017ost}
Jana, S.; Chakravarty, G.K.; Mohanty, S.
\newblock {Constraints on Born-Infeld gravity from the speed of gravitational
  waves after GW170817 and GRB 170817A}.
\newblock {\em Phys. Rev. D} {\bf 2018}, {\em 97},~084011,
  \href{http://xxx.lanl.gov/abs/1711.04137}{{\normalfont
  [arXiv:gr-qc/1711.04137]}}.
\newblock
  doi:{\changeurlcolor{black}\href{https://doi.org/10.1103/PhysRevD.97.084011}{\detokenize{10.1103/PhysRevD.97.084011}}}.

\bibitem[Barragan and Olmo(2010)]{Barragan:2010qb}
Barragan, C.; Olmo, G.J.
\newblock {Isotropic and Anisotropic Bouncing Cosmologies in Palatini Gravity}.
\newblock {\em Phys. Rev.} {\bf 2010}, {\em D82},~084015,
  \href{http://xxx.lanl.gov/abs/1005.4136}{{\normalfont
  [arXiv:gr-qc/1005.4136]}}.
\newblock
  doi:{\changeurlcolor{black}\href{https://doi.org/10.1103/PhysRevD.82.084015}{\detokenize{10.1103/PhysRevD.82.084015}}}.

\bibitem[Afonso \em{et~al.}(2017)Afonso, Olmo, and
  Rubiera-Garcia]{Afonso:2017aci}
Afonso, V.I.; Olmo, G.J.; Rubiera-Garcia, D.
\newblock {Scalar geons in Born-Infeld gravity}.
\newblock {\em JCAP} {\bf 2017}, {\em 08},~031,
  \href{http://xxx.lanl.gov/abs/1705.01065}{{\normalfont
  [arXiv:gr-qc/1705.01065]}}.
\newblock
  doi:{\changeurlcolor{black}\href{https://doi.org/10.1088/1475-7516/2017/08/031}{\detokenize{10.1088/1475-7516/2017/08/031}}}.

\bibitem[Solà \em{et~al.}(2017)Solà, Gómez-Valent, and
  de~Cruz~Pérez]{Sola:2016jky}
Solà, J.; Gómez-Valent, A.; de~Cruz~Pérez, J.
\newblock {First evidence of running cosmic vacuum: challenging the concordance
  model}.
\newblock {\em Astrophys. J.} {\bf 2017}, {\em 836},~43,
  \href{http://xxx.lanl.gov/abs/1602.02103}{{\normalfont
  [arXiv:astro-ph.CO/1602.02103]}}.
\newblock
  doi:{\changeurlcolor{black}\href{https://doi.org/10.3847/1538-4357/836/1/43}{\detokenize{10.3847/1538-4357/836/1/43}}}.

\bibitem[Barrow \em{et~al.}(2021)Barrow, Basilakos, and
  Saridakis]{Barrow:2020kug}
Barrow, J.D.; Basilakos, S.; Saridakis, E.N.
\newblock {Big Bang Nucleosynthesis constraints on Barrow entropy}.
\newblock {\em Phys. Lett.} {\bf 2021}, {\em B815},~136134,
  \href{http://xxx.lanl.gov/abs/2010.00986}{{\normalfont
  [arXiv:gr-qc/2010.00986]}}.
\newblock
  doi:{\changeurlcolor{black}\href{https://doi.org/10.1016/j.physletb.2021.136134}{\detokenize{10.1016/j.physletb.2021.136134}}}.

\bibitem[Jim\'enez \em{et~al.}(2021)Jim\'enez, Delhom, Olmo, and
  Orazi]{Jimenez:2021sqd}
Jim\'enez, J.B.; Delhom, A.; Olmo, G.J.; Orazi, E.
\newblock {Born-Infeld gravity: Constraints from light-by-light scattering and
  an effective field theory perspective}.
\newblock {\em Phys. Lett. B} {\bf 2021}, {\em 820},~136479,
  \href{http://xxx.lanl.gov/abs/2104.01647}{{\normalfont
  [arXiv:gr-qc/2104.01647]}}.
\newblock
  doi:{\changeurlcolor{black}\href{https://doi.org/10.1016/j.physletb.2021.136479}{\detokenize{10.1016/j.physletb.2021.136479}}}.

\bibitem[Latorre \em{et~al.}(2018)Latorre, Olmo, and Ronco]{Latorre:2017uve}
Latorre, A.D.I.; Olmo, G.J.; Ronco, M.
\newblock {Observable traces of non-metricity: new constraints on metric-affine
  gravity}.
\newblock {\em Phys. Lett.} {\bf 2018}, {\em B780},~294--299,
  \href{http://xxx.lanl.gov/abs/1709.04249}{{\normalfont
  [arXiv:hep-th/1709.04249]}}.
\newblock
  doi:{\changeurlcolor{black}\href{https://doi.org/10.1016/j.physletb.2018.03.002}{\detokenize{10.1016/j.physletb.2018.03.002}}}.

\bibitem[Delhom \em{et~al.}(2020)Delhom, Miralles, and
  Peñuelas]{Delhom:2019wir}
Delhom, A.; Miralles, V.; Peñuelas, A.
\newblock {Effective interactions in Ricci-Based Gravity below the
  non-metricity scale}.
\newblock {\em Eur. Phys. J.} {\bf 2020}, {\em C80},~340,
  \href{http://xxx.lanl.gov/abs/1907.05615}{{\normalfont
  [arXiv:hep-th/1907.05615]}}.
\newblock
  doi:{\changeurlcolor{black}\href{https://doi.org/10.1140/epjc/s10052-020-7880-3}{\detokenize{10.1140/epjc/s10052-020-7880-3}}}.

\bibitem[Jimenez and Loeb(2002)]{Jimenez:2001gg}
Jimenez, R.; Loeb, A.
\newblock {Constraining cosmological parameters based on relative galaxy ages}.
\newblock {\em Astrophys. J.} {\bf 2002}, {\em 573},~37--42,
  \href{http://xxx.lanl.gov/abs/astro-ph/0106145}{{\normalfont
  [arXiv:astro-ph/astro-ph/0106145]}}.
\newblock
  doi:{\changeurlcolor{black}\href{https://doi.org/10.1086/340549}{\detokenize{10.1086/340549}}}.

\bibitem[Moresco \em{et~al.}(2012{\natexlab{a}})Moresco, Verde, Pozzetti,
  Jimenez, and Cimatti]{Moresco:2012by}
Moresco, M.; Verde, L.; Pozzetti, L.; Jimenez, R.; Cimatti, A.
\newblock {New constraints on cosmological parameters and neutrino properties
  using the expansion rate of the Universe to z~1.75}.
\newblock {\em JCAP} {\bf 2012}, {\em 1207},~053,
  \href{http://xxx.lanl.gov/abs/1201.6658}{{\normalfont
  [arXiv:astro-ph.CO/1201.6658]}}.
\newblock
  doi:{\changeurlcolor{black}\href{https://doi.org/10.1088/1475-7516/2012/07/053}{\detokenize{10.1088/1475-7516/2012/07/053}}}.

\bibitem[Moresco \em{et~al.}(2012{\natexlab{b}})Moresco et~al.]{Moresco:2012jh}
Moresco, M.; others.
\newblock {Improved constraints on the expansion rate of the Universe up to
  z~1.1 from the spectroscopic evolution of cosmic chronometers}.
\newblock {\em JCAP} {\bf 2012}, {\em 1208},~006,
  \href{http://xxx.lanl.gov/abs/1201.3609}{{\normalfont
  [arXiv:astro-ph.CO/1201.3609]}}.
\newblock
  doi:{\changeurlcolor{black}\href{https://doi.org/10.1088/1475-7516/2012/08/006}{\detokenize{10.1088/1475-7516/2012/08/006}}}.

\bibitem[Moresco(2015)]{Moresco:2015cya}
Moresco, M.
\newblock {Raising the bar: new constraints on the Hubble parameter with cosmic
  chronometers at z ~ 2}.
\newblock {\em Mon. Not. Roy. Astron. Soc.} {\bf 2015}, {\em 450},~L16--L20,
  \href{http://xxx.lanl.gov/abs/1503.01116}{{\normalfont
  [arXiv:astro-ph.CO/1503.01116]}}.
\newblock
  doi:{\changeurlcolor{black}\href{https://doi.org/10.1093/mnrasl/slv037}{\detokenize{10.1093/mnrasl/slv037}}}.

\bibitem[Moresco \em{et~al.}(2016)Moresco, Pozzetti, Cimatti, Jimenez,
  Maraston, Verde, Thomas, Citro, Tojeiro, and Wilkinson]{Moresco:2016mzx}
Moresco, M.; Pozzetti, L.; Cimatti, A.; Jimenez, R.; Maraston, C.; Verde, L.;
  Thomas, D.; Citro, A.; Tojeiro, R.; Wilkinson, D.
\newblock {A 6\% measurement of the Hubble parameter at $z\sim0.45$: direct
  evidence of the epoch of cosmic re-acceleration}.
\newblock {\em JCAP} {\bf 2016}, {\em 1605},~014,
  \href{http://xxx.lanl.gov/abs/1601.01701}{{\normalfont
  [arXiv:astro-ph.CO/1601.01701]}}.
\newblock
  doi:{\changeurlcolor{black}\href{https://doi.org/10.1088/1475-7516/2016/05/014}{\detokenize{10.1088/1475-7516/2016/05/014}}}.

\bibitem[Scolnic \em{et~al.}(2018)Scolnic et~al.]{Scolnic:2017caz}
Scolnic, D.; others.
\newblock {The Complete Light-curve Sample of Spectroscopically Confirmed SNe
  Ia from Pan-STARRS1 and Cosmological Constraints from the Combined Pantheon
  Sample}.
\newblock {\em Astrophys. J.} {\bf 2018}, {\em 859},~101,
  \href{http://xxx.lanl.gov/abs/1710.00845}{{\normalfont
  [arXiv:astro-ph.CO/1710.00845]}}.
\newblock
  doi:{\changeurlcolor{black}\href{https://doi.org/10.3847/1538-4357/aab9bb}{\detokenize{10.3847/1538-4357/aab9bb}}}.

\bibitem[Anagnostopoulos \em{et~al.}(2020)Anagnostopoulos, Basilakos, and
  Saridakis]{Anagnostopoulos:2020ctz}
Anagnostopoulos, F.K.; Basilakos, S.; Saridakis, E.N.
\newblock {Observational constraints on Barrow holographic dark energy}.
\newblock {\em Eur. Phys. J.} {\bf 2020}, {\em C80},~826,
  \href{http://xxx.lanl.gov/abs/2005.10302}{{\normalfont
  [arXiv:gr-qc/2005.10302]}}.
\newblock
  doi:{\changeurlcolor{black}\href{https://doi.org/10.1140/epjc/s10052-020-8360-5}{\detokenize{10.1140/epjc/s10052-020-8360-5}}}.

\bibitem[Roberts \em{et~al.}(2017)Roberts, Horne, Hodson, and
  Leggat]{Roberts:2017nkm}
Roberts, C.; Horne, K.; Hodson, A.O.; Leggat, A.D.
\newblock {Tests of $\Lambda$CDM and Conformal Gravity using GRB and Quasars as
  Standard Candles out to $z \sim 8$} {\bf 2017}.
\newblock  \href{http://xxx.lanl.gov/abs/1711.10369}{{\normalfont
  [arXiv:astro-ph.CO/1711.10369]}}.

\bibitem[Demianski \em{et~al.}(2017)Demianski, Piedipalumbo, Sawant, and
  Amati]{Demianski:2016zxi}
Demianski, M.; Piedipalumbo, E.; Sawant, D.; Amati, L.
\newblock {Cosmology with gamma-ray bursts: I. The Hubble diagram through the
  calibrated $E_{\rm p,i}$ - $E_{\rm iso}$ correlation}.
\newblock {\em Astron. Astrophys.} {\bf 2017}, {\em 598},~A112,
  \href{http://xxx.lanl.gov/abs/1610.00854}{{\normalfont
  [arXiv:astro-ph.CO/1610.00854]}}.
\newblock
  doi:{\changeurlcolor{black}\href{https://doi.org/10.1051/0004-6361/201628909}{\detokenize{10.1051/0004-6361/201628909}}}.

\bibitem[Hogg \em{et~al.}(2020)Hogg, Martinelli, and Nesseris]{Hogg:2020ktc}
Hogg, N.B.; Martinelli, M.; Nesseris, S.
\newblock {Constraints on the distance duality relation with standard sirens}
  {\bf 2020}.
\newblock  \href{http://xxx.lanl.gov/abs/2007.14335}{{\normalfont
  [arXiv:astro-ph.CO/2007.14335]}}.

\bibitem[Martinelli \em{et~al.}(2020)Martinelli et~al.]{Martinelli:2020hud}
Martinelli, M.; others.
\newblock {Euclid: Forecast constraints on the cosmic distance duality relation
  with complementary external probes} {\bf 2020}.
\newblock  \href{http://xxx.lanl.gov/abs/2007.16153}{{\normalfont
  [arXiv:astro-ph.CO/2007.16153]}}.

\bibitem[Benisty and Staicova(2020)]{Benisty:2020otr}
Benisty, D.; Staicova, D.
\newblock {Testing Low-Redshift Cosmic Acceleration with the Complete Baryon
  Acoustic Oscillations data collection} {\bf 2020}.
\newblock  \href{http://xxx.lanl.gov/abs/2009.10701}{{\normalfont
  [arXiv:astro-ph.CO/2009.10701]}}.

\bibitem[Percival \em{et~al.}(2010)Percival et~al.]{Percival:2009xn}
Percival, W.J.; others.
\newblock {Baryon Acoustic Oscillations in the Sloan Digital Sky Survey Data
  Release 7 Galaxy Sample}.
\newblock {\em Mon. Not. Roy. Astron. Soc.} {\bf 2010}, {\em 401},~2148--2168,
  \href{http://xxx.lanl.gov/abs/0907.1660}{{\normalfont
  [arXiv:astro-ph.CO/0907.1660]}}.
\newblock
  doi:{\changeurlcolor{black}\href{https://doi.org/10.1111/j.1365-2966.2009.15812.x}{\detokenize{10.1111/j.1365-2966.2009.15812.x}}}.

\bibitem[Beutler \em{et~al.}(2011)Beutler, Blake, Colless, Jones,
  Staveley-Smith, Campbell, Parker, Saunders, and Watson]{Beutler:2011hx}
Beutler, F.; Blake, C.; Colless, M.; Jones, D.H.; Staveley-Smith, L.; Campbell,
  L.; Parker, Q.; Saunders, W.; Watson, F.
\newblock {The 6dF Galaxy Survey: Baryon Acoustic Oscillations and the Local
  Hubble Constant}.
\newblock {\em Mon. Not. Roy. Astron. Soc.} {\bf 2011}, {\em 416},~3017--3032,
  \href{http://xxx.lanl.gov/abs/1106.3366}{{\normalfont
  [arXiv:astro-ph.CO/1106.3366]}}.
\newblock
  doi:{\changeurlcolor{black}\href{https://doi.org/10.1111/j.1365-2966.2011.19250.x}{\detokenize{10.1111/j.1365-2966.2011.19250.x}}}.

\bibitem[Busca \em{et~al.}(2013)Busca et~al.]{Busca:2012bu}
Busca, N.G.; others.
\newblock {Baryon Acoustic Oscillations in the Ly-$\alpha$ forest of BOSS
  quasars}.
\newblock {\em Astron. Astrophys.} {\bf 2013}, {\em 552},~A96,
  \href{http://xxx.lanl.gov/abs/1211.2616}{{\normalfont
  [arXiv:astro-ph.CO/1211.2616]}}.
\newblock
  doi:{\changeurlcolor{black}\href{https://doi.org/10.1051/0004-6361/201220724}{\detokenize{10.1051/0004-6361/201220724}}}.

\bibitem[Anderson \em{et~al.}(2013)Anderson et~al.]{Anderson:2012sa}
Anderson, L.; others.
\newblock {The clustering of galaxies in the SDSS-III Baryon Oscillation
  Spectroscopic Survey: Baryon Acoustic Oscillations in the Data Release 9
  Spectroscopic Galaxy Sample}.
\newblock {\em Mon. Not. Roy. Astron. Soc.} {\bf 2013}, {\em 427},~3435--3467,
  \href{http://xxx.lanl.gov/abs/1203.6594}{{\normalfont
  [arXiv:astro-ph.CO/1203.6594]}}.
\newblock
  doi:{\changeurlcolor{black}\href{https://doi.org/10.1111/j.1365-2966.2012.22066.x}{\detokenize{10.1111/j.1365-2966.2012.22066.x}}}.

\bibitem[Seo \em{et~al.}(2012)Seo et~al.]{Seo:2012xy}
Seo, H.J.; others.
\newblock {Acoustic scale from the angular power spectra of SDSS-III DR8
  photometric luminous galaxies}.
\newblock {\em Astrophys. J.} {\bf 2012}, {\em 761},~13,
  \href{http://xxx.lanl.gov/abs/1201.2172}{{\normalfont
  [arXiv:astro-ph.CO/1201.2172]}}.
\newblock
  doi:{\changeurlcolor{black}\href{https://doi.org/10.1088/0004-637X/761/1/13}{\detokenize{10.1088/0004-637X/761/1/13}}}.

\bibitem[Ross \em{et~al.}(2015)Ross, Samushia, Howlett, Percival, Burden, and
  Manera]{Ross:2014qpa}
Ross, A.J.; Samushia, L.; Howlett, C.; Percival, W.J.; Burden, A.; Manera, M.
\newblock {The clustering of the SDSS DR7 main Galaxy sample – I. A 4 per
  cent distance measure at $z = 0.15$}.
\newblock {\em Mon. Not. Roy. Astron. Soc.} {\bf 2015}, {\em 449},~835--847,
  \href{http://xxx.lanl.gov/abs/1409.3242}{{\normalfont
  [arXiv:astro-ph.CO/1409.3242]}}.
\newblock
  doi:{\changeurlcolor{black}\href{https://doi.org/10.1093/mnras/stv154}{\detokenize{10.1093/mnras/stv154}}}.

\bibitem[Tojeiro \em{et~al.}(2014)Tojeiro et~al.]{Tojeiro:2014eea}
Tojeiro, R.; others.
\newblock {The clustering of galaxies in the SDSS-III Baryon Oscillation
  Spectroscopic Survey: galaxy clustering measurements in the low redshift
  sample of Data Release 11}.
\newblock {\em Mon. Not. Roy. Astron. Soc.} {\bf 2014}, {\em 440},~2222--2237,
  \href{http://xxx.lanl.gov/abs/1401.1768}{{\normalfont
  [arXiv:astro-ph.CO/1401.1768]}}.
\newblock
  doi:{\changeurlcolor{black}\href{https://doi.org/10.1093/mnras/stu371}{\detokenize{10.1093/mnras/stu371}}}.

\bibitem[Bautista \em{et~al.}(2018)Bautista et~al.]{Bautista:2017wwp}
Bautista, J.E.; others.
\newblock {The SDSS-IV extended Baryon Oscillation Spectroscopic Survey: Baryon
  Acoustic Oscillations at redshift of 0.72 with the DR14 Luminous Red Galaxy
  Sample}.
\newblock {\em Astrophys. J.} {\bf 2018}, {\em 863},~110,
  \href{http://xxx.lanl.gov/abs/1712.08064}{{\normalfont
  [arXiv:astro-ph.CO/1712.08064]}}.
\newblock
  doi:{\changeurlcolor{black}\href{https://doi.org/10.3847/1538-4357/aacea5}{\detokenize{10.3847/1538-4357/aacea5}}}.

\bibitem[de~Carvalho \em{et~al.}(2018)de~Carvalho, Bernui, Carvalho, Novaes,
  and Xavier]{deCarvalho:2017xye}
de~Carvalho, E.; Bernui, A.; Carvalho, G.C.; Novaes, C.P.; Xavier, H.S.
\newblock {Angular Baryon Acoustic Oscillation measure at $z=2.225$ from the
  SDSS quasar survey}.
\newblock {\em JCAP} {\bf 2018}, {\em 1804},~064,
  \href{http://xxx.lanl.gov/abs/1709.00113}{{\normalfont
  [arXiv:astro-ph.CO/1709.00113]}}.
\newblock
  doi:{\changeurlcolor{black}\href{https://doi.org/10.1088/1475-7516/2018/04/064}{\detokenize{10.1088/1475-7516/2018/04/064}}}.

\bibitem[Ata \em{et~al.}(2018)Ata et~al.]{Ata:2017dya}
Ata, M.; others.
\newblock {The clustering of the SDSS-IV extended Baryon Oscillation
  Spectroscopic Survey DR14 quasar sample: first measurement of baryon acoustic
  oscillations between redshift 0.8 and 2.2}.
\newblock {\em Mon. Not. Roy. Astron. Soc.} {\bf 2018}, {\em 473},~4773--4794,
  \href{http://xxx.lanl.gov/abs/1705.06373}{{\normalfont
  [arXiv:astro-ph.CO/1705.06373]}}.
\newblock
  doi:{\changeurlcolor{black}\href{https://doi.org/10.1093/mnras/stx2630}{\detokenize{10.1093/mnras/stx2630}}}.

\bibitem[Abbott \em{et~al.}(2019)Abbott et~al.]{Abbott:2017wcz}
Abbott, T.M.C.; others.
\newblock {Dark Energy Survey Year 1 Results: Measurement of the Baryon
  Acoustic Oscillation scale in the distribution of galaxies to redshift 1}.
\newblock {\em Mon. Not. Roy. Astron. Soc.} {\bf 2019}, {\em 483},~4866--4883,
  \href{http://xxx.lanl.gov/abs/1712.06209}{{\normalfont
  [arXiv:astro-ph.CO/1712.06209]}}.
\newblock
  doi:{\changeurlcolor{black}\href{https://doi.org/10.1093/mnras/sty3351}{\detokenize{10.1093/mnras/sty3351}}}.

\bibitem[Molavi and Khodam-Mohammadi(2019)]{Molavi:2019mlh}
Molavi, Z.; Khodam-Mohammadi, A.
\newblock {Observational tests of Gauss-Bonnet like dark energy model}.
\newblock {\em Eur. Phys. J. Plus} {\bf 2019}, {\em 134},~254,
  \href{http://xxx.lanl.gov/abs/1906.05668}{{\normalfont
  [arXiv:gr-qc/1906.05668]}}.
\newblock
  doi:{\changeurlcolor{black}\href{https://doi.org/10.1140/epjp/i2019-12723-x}{\detokenize{10.1140/epjp/i2019-12723-x}}}.

\bibitem[Riess \em{et~al.}(2021)Riess, Casertano, Yuan, Bowers, Macri, Zinn,
  and Scolnic]{Riess:2020fzl}
Riess, A.G.; Casertano, S.; Yuan, W.; Bowers, J.B.; Macri, L.; Zinn, J.C.;
  Scolnic, D.
\newblock {Cosmic Distances Calibrated to 1\% Precision with Gaia EDR3
  Parallaxes and Hubble Space Telescope Photometry of 75 Milky Way Cepheids
  Confirm Tension with $\Lambda$CDM}.
\newblock {\em Astrophys. J. Lett.} {\bf 2021}, {\em 908},~L6,
  \href{http://xxx.lanl.gov/abs/2012.08534}{{\normalfont
  [arXiv:astro-ph.CO/2012.08534]}}.
\newblock
  doi:{\changeurlcolor{black}\href{https://doi.org/10.3847/2041-8213/abdbaf}{\detokenize{10.3847/2041-8213/abdbaf}}}.

\bibitem[Foreman-Mackey \em{et~al.}(2013)Foreman-Mackey, Hogg, Lang, and
  Goodman]{ForemanMackey:2012ig}
Foreman-Mackey, D.; Hogg, D.W.; Lang, D.; Goodman, J.
\newblock {emcee: The MCMC Hammer}.
\newblock {\em Publ. Astron. Soc. Pac.} {\bf 2013}, {\em 125},~306--312,
  \href{http://xxx.lanl.gov/abs/1202.3665}{{\normalfont
  [arXiv:astro-ph.IM/1202.3665]}}.
\newblock
  doi:{\changeurlcolor{black}\href{https://doi.org/10.1086/670067}{\detokenize{10.1086/670067}}}.

\bibitem[Handley \em{et~al.}(2015)Handley, Hobson, and
  Lasenby]{Handley:2015fda}
Handley, W.J.; Hobson, M.P.; Lasenby, A.N.
\newblock {PolyChord: nested sampling for cosmology}.
\newblock {\em Mon. Not. Roy. Astron. Soc.} {\bf 2015}, {\em 450},~L61--L65,
  \href{http://xxx.lanl.gov/abs/1502.01856}{{\normalfont
  [arXiv:astro-ph.CO/1502.01856]}}.
\newblock
  doi:{\changeurlcolor{black}\href{https://doi.org/10.1093/mnrasl/slv047}{\detokenize{10.1093/mnrasl/slv047}}}.

\bibitem[Lewis(2019)]{Lewis:2019xzd}
Lewis, A.
\newblock {GetDist: a Python package for analysing Monte Carlo samples} {\bf
  2019}.
\newblock  \href{http://xxx.lanl.gov/abs/1910.13970}{{\normalfont
  [arXiv:astro-ph.IM/1910.13970]}}.

\bibitem[Beltran~Jimenez \em{et~al.}(2017)Beltran~Jimenez, Heisenberg, Olmo,
  and Rubiera-Garcia]{BeltranJimenez:2017uwv}
Beltran~Jimenez, J.; Heisenberg, L.; Olmo, G.J.; Rubiera-Garcia, D.
\newblock {On gravitational waves in Born-Infeld inspired non-singular
  cosmologies}.
\newblock {\em JCAP} {\bf 2017}, {\em 10},~029,
  \href{http://xxx.lanl.gov/abs/1707.08953}{{\normalfont
  [arXiv:hep-th/1707.08953]}}.
\newblock [Erratum: JCAP 08, E01 (2018)],
  doi:{\changeurlcolor{black}\href{https://doi.org/10.1088/1475-7516/2017/10/029}{\detokenize{10.1088/1475-7516/2017/10/029}}}.

\bibitem[Escamilla-Rivera \em{et~al.}(2012)Escamilla-Rivera, Banados, and
  Ferreira]{EscamillaRivera:2012vz}
Escamilla-Rivera, C.; Banados, M.; Ferreira, P.G.
\newblock {A tensor instability in the Eddington inspired Born-Infeld Theory of
  Gravity}.
\newblock {\em Phys. Rev. D} {\bf 2012}, {\em 85},~087302,
  \href{http://xxx.lanl.gov/abs/1204.1691}{{\normalfont
  [arXiv:gr-qc/1204.1691]}}.
\newblock
  doi:{\changeurlcolor{black}\href{https://doi.org/10.1103/PhysRevD.85.087302}{\detokenize{10.1103/PhysRevD.85.087302}}}.

\end{thebibliography}

\end{document}